\documentclass[useAMS,usenatbib,usegraphicx,twocolumn]{mn2e}
  
\usepackage{aas_macros}
\usepackage{threeparttable}
\usepackage{grffile}
\usepackage{amssymb,amsmath}
\usepackage{multirow}
\usepackage{wallpaper}
\usepackage{mathbbol}
\usepackage{rotating}
\usepackage{blindtext}

\usepackage{tocloft}
\newcommand{\fat}{\textbf}
\newcommand{\ita}{\textit}
\newcommand{\del}{\partial}

\newcommand{\beq}{\begin{equation}}
\newcommand{\eeq}{\end{equation}}  
\newcommand{\RNum}[1]{\uppercase\expandafter{\romannumeral #1\relax}}

\newcommand{\pc}{\,\mathrm{pc}}
\newcommand{\myr}{\,\mathrm{Myr}}
\defcitealias{Koertgen15}{\scshape KB15}
\defcitealias{Koertgen16}{\scshape K16}
\title[The Density-PDF in MCs]{On the shape and completeness of the column density probability distribution function of molecular clouds}
  \author[B. K\"ortgen]
  {Bastian~K\"ortgen$^{1}$\thanks{bkoertgen@hs.uni-hamburg.de}, Christoph~Federrath$^{2}$ and Robi~Banerjee$^{1}$ \\
  $^{1}$ Hamburger Sternwarte, Universit\"at Hamburg, Gojenbergsweg 112, 21029 Hamburg, Germany \\
  $^{2}$ Research School of Astronomy and Astrophysics, Australian National University, Canberra, ACT 2611, Australia\\
  }

\date{Released 2018}

\pagerange{\pageref{firstpage}--\pageref{lastpage}} \pubyear{2018}

\begin{document}

\label{firstpage}
\maketitle

\begin{abstract}
Both observational and theoretical research over the past decade has demonstrated that the probability distribution function (PDF) of the gas density in turbulent molecular clouds is a key ingredient for understanding star formation.
It has recently been argued that the PDF of molecular clouds is a pure power-law distribution. It has been claimed that the log-normal part is ruled out when using only the 
part of the PDF up/down to which it is complete, that is where the column density contours are still closed. By using the results from high-resolution magnetohydrodynamical simulations of 
molecular cloud formation and evolution, we find that the column density PDF is indeed composed of a log-normal and, 
if including self-gravity, a power-law part. We show that insufficient sampling of a molecular cloud results in closed contours that cut off the log-normal part. In contrast, systematically increasing the field of view and sampling the entire cloud yields a completeness limit 
at the lower column densities, which also recovers the log-normal part. This demonstrates that the field of view must be sufficiently large for the PDF to be complete down to its log-normal part, which has important
 implications for predictions of star-formation activity based on the PDF.
\end{abstract}

\section{Introduction}
The (column-) density probability distribution function (henceforth N-PDF or PDF) has become a powerful tool to analyse the dynamics of molecular clouds both from 
an observational \citep{Elmegreen04,Kainulainen09,Brunt10,Ginsburg13,Kainulainen13b,Lombardi14,Burkhart15,Schneider15a,Schneider16,Federrath16d} and a theoretical 
perspective \citep{Passot98,Federrath08,Federrath10b,Konstandin12,Girichidis14,Nolan15,Federrath15,Burkhart17}. For isothermal gas, its shape is best described by a log-normal function as a consequence of interacting turbulent shocks 
modifying the density structure \citep{Vazquez94}. Mathematically, this shape can be understood from the central-limit theorem of an ensemble of independent events \citep{Passot98,Kritsuk07,Federrath10b}. In contrast, in a thermally unstable gas, a multiphase medium is formed, which consequently results 
in a multi-peaked PDF, with the peaks being located at the characteristic density of each phase \citep{Pikelner68,Field69,Gazol01,Gazol05}. However, 
the clear separation of the different peaks might become washed out, when the turbulent motions in the unstable gas become sufficiently 
strong \citep{Vazquez00,Audit05}. \\
The high-density fraction of the PDF has either a log-normal shape for isothermal, non-self-gravitating gas or forms a power-law tail, when the dense gas structures are dominated by gravity \citep{Federrath13,Kainulainen14,Girichidis14} or have an equation of state softer than isothermal (i.e. $\gamma<1$). A physical explanation for the emergence of a power-law tail due to gravity was given 
by \citet{Kritsuk11a}. The authors showed that the formation of power-law density profiles in (small-scale) gravitationally collapsing regions naturally leads to 
a power-law in the density PDF and thus also in the N-PDF \citep[see also][and references therein]{Federrath13}. In contrast, 
\citet{Lombardi15} suggest that the power-law form of the PDF arises because the entire molecular cloud can be described 
by a power-law density profile.\\
 But, there can also be other reasons for a power-law tail. It was shown that gas compression by 
external pressure \citep{Tremblin14} or sufficiently strong turbulence in the thermally bistable gas can also yield a power-law tail 
\citep{Vazquez00}. \citet{Ballesteros11b} have studied the evolution of the N-PDF of a molecular cloud formed in 
numerical simulations of a thermally unstable gas including self-gravity. They showed that a power-law tail naturally develops when 
self-gravity becomes important, but that the double-peak signature arising from thermal instability is still visible in the PDF. With 
these previous studies in mind, there might indeed arise a PDF, where the signature of 
thermal instability is blurred and which thus shows a power-law tail all the way down to the lowest densities.\\
The information content of the N-PDF is biased by several effects, such as by contamination from clouds lying in the fore- or background \citep{Lombardi15,Schneider15b,Ossenkopf16}. As discussed in 
\citet{Schneider15b} and \citet{Ossenkopf16}, this can be corrected for by subtraction of a constant offset. However, as \citet{Schneider15b} point out, such an approach might over- or underestimate the 
influence of the line-of-sight contamination. In addition, \citet{Lombardi15} and \citet{Ossenkopf16} show that it is mainly the 
low-density part of the PDF that is affected by the contamination, whereas the high-density (power-law) part is nearly unaffected.\\
 However, it has recently been argued that N-PDFs, after subtraction of an offset to 
 correct for contamination, are no longer log-normal, but rather show only a power-law form \citep{Lombardi15}. A theoretical approach was conducted by \citet{Ward14}, who study N-PDFs in synthetic observations
 of simulated molecular clouds. The authors showed that dynamically old clouds show a log-normal part in combination with a power-law tail, but that the log-normal part systematically falls below the 
 visual extinction threshold. \citet{Ward14} conclude that a pure power-law distribution might thus be due only to limited observational resolution.\\
 The retrieved information of a PDF is only reliable within its completeness limit \citep{Kainulainen13b}. From an observational perspective, it has been suggested that the PDF is complete down to the smallest column density with a closed 
 contour \citep{Kainulainen13b}. By using this definition, it has recently been argued that column density PDFs 
of a variety of high galactic latitude clouds (diffuse or star-forming) have a power-law shape when the last closed contour defines the completeness of the PDF \citep{Alves17}. These authors stated that there is no observational evidence for log-normal PDFs and that molecular clouds have PDFs well described by power laws.\\
In this study, we analyse the N-PDF of molecular clouds formed in three-dimensional, magnetohydrodynamical simulations of converging flows. We estimate the value of the \ita{last closed contour} 
as a function of a varying field of view for a star-forming and a quiescent region within the formed cloud complex, as well as for the entire 
complex and study the obtained N-PDF. We show that the value of the last closed contour moves towards higher column densities for a decreasing field of view. We find that the 
log-normal part of the N-PDF is well within the completeness of the PDF for a sufficiently large field of view, emphasising that N-PDFs are indeed composed of a log-normal part (and a power-law tail, when gravitational collapse sets in).\\
This paper is organised as follows: In section~\ref{flash} we introduce the numerical setup and the initial conditions. Section~\ref{flows} provides a brief description of the general time evolution of the 
formed clouds and a comparison of the resulting N-PDF for a case with and without self-gravity but otherwise identical parameters. This is followed by the presentation of our results of the N-PDF for different extents of the fields of view of two molecular cloud regions in section~\ref{results}, and our conclusions are given in section~\ref{conclusions}.

\section{Simulation data and methods}\label{flash}
The simulations presented here were carried out with the \textsc{flash} code in version 2.5 \citep{FLASH00}.\\
We set up two cylindrical flows of warm neutral medium (WNM) gas with a length $l=112\pc$ and a radius of $R=64\pc$, which collide head-on in the centre of the cubic simulation domain, which has a volume of 
$V_\mathrm{sim}=(256\pc)^3$. The computational domain is initially filled with gas of number density $n=1\,\mathrm{cm}^{-3}$ and temperature $T=5000\,\mathrm{K}$. The gas is able to heat and cool via optically-thin 
radiation, which is given in tabulated form according to the prescription by \citet{Koyama02} and used as a source term in the energy equation. The fitting functions for the heating and cooling rates are given by 
\beq 
\Gamma = 2\times10^{-26}\,\mathrm{erg s}^{-1},
\eeq
with the heating rate $\Gamma$, and 
\beq
\begin{split}
\frac{\Lambda(T)}{\Gamma} &=10^7\mathrm{exp}\left(\frac{-1.184\times10^5}{T+1000}\right)\\
&\quad +1.4\times10^{-2}\sqrt{T}\mathrm{exp}\left(\frac{-92}{T}\right),
\end{split}
\eeq
where $\Lambda(T)$ is the temperature-dependent cooling rate and $T$ the temperature in Kelvin \citep{Koyama02,Vazquez07}.
By using this definition, the gas is initially in the thermally 
unstable regime and will develop into a two-phase medium.\\
The sound speed at the initial temperature is \mbox{$c_\mathrm{S}=5.7\,\mathrm{km\,s}^{-1}$} and the WNM flows are initialised with a flow velocity of $v_\mathrm{F}=11.4\,\mathrm{km\,s}^{-1}$, 
which corresponds to a isothermal Mach number of $\mathcal{M}_\mathrm{F}=2$, being thus mildly supersonic. In order to trigger dynamical and thermal instabilities, the flows are additionally turbulent with 
$\mathcal{M}_\mathrm{RMS}=1$, and the energy spectrum is of Burgers type, $E(k)\propto k^{-2}$ \citep{Ossenkopf02,Heyer04}.\\
Since the interstellar medium of galaxies is also magnetised \citep{Crutcher10,Beck12}, we add a magnetic field $\fat{B}=B_0\hat{\fat{x}}$ 
aligned parallel to the flows with $\hat{\fat{x}}$ being the unit vector in the x-direction. The initial magnitude is $B_0=3\,\mu\mathrm{G}$ in all simulations, indicating that the flows are magnetically critical 
with a normalised mass-to-magnetic flux ratio of $\mu/\mu_\mathrm{crit}\sim1$. Note that accretion of gas from the environment will rapidly increase the mass-to-flux ratio so that the cloud will be highly
supercritical\footnote{Please note that this statement is only true for the region defined by the flow geometry. The mass-to-flux ratio of the whole box stays constant (and 
supercritical) due to 
the choice of periodic boundary conditions.}.\\
The simulations presented here use the adaptive mesh refinement technique \citep{Berger84,Berger89}. The root grid is at resolution of 
\mbox{$\Delta x_\mathrm{root}=64\,\mathrm{pc}$}. To sufficiently resolve the dynamics at the initial stages, the collision layer is refined to a resolution of 
\mbox{$\Delta x_\mathrm{layer}=1\,\mathrm{pc}$}. We allow for a maximum of 12 refinement levels, which gives a minimum cell size of 
\mbox{$\Delta x=0.0156\pc$}. The grid is refined once the local Jeans length is 
resolved with less than 8 grid cells. To prevent the 
gas from fragmenting artificially, we introduce Lagrangian sink particles when the gas is, besides other checks, at a density of $n_\mathrm{thresh}=3\times10^5\,\mathrm{cm}^{-3}$, which represent stellar clusters 
rather than individual stars due to the limited spatial resolution. This ensures that the simulations are fully resolved and fragmentation of the cloud is due to physical processes \citep{Truelove97,Federrath10}. 
Feedback from sink particles via winds, radiation or supernovae is not included in the simulations, such that our simulations are probing the density PDF produced by turbulent, magnetised inflowing gas streams and gravity alone.\\
With these initial conditions at hand, the following set of equations of ideal magnetohydrodynamics in combination with Poisson's equation for the self-gravity of the gas and heating and cooling is solved during each 
time-step
\beq
\begin{split}
 &\frac{\del}{\del t}\varrho + \nabla\cdot\left(\varrho\fat{u}\right)=0\\
 &\varrho\frac{\del}{\del t}\fat{u}+\varrho\left(\fat{u}\cdot\nabla\right)\fat{u}=-\nabla P_\mathrm{tot}+\varrho\fat{g}+\frac{\left(\fat{B}\cdot\nabla\right)\fat{B}}{4\pi}\\
 &\frac{\del}{\del t}E+\nabla\cdot\left[\left(E+P_\mathrm{tot}\right)\fat{u}-\frac{\left(\fat{B}\cdot\fat{u}\right)\fat{B}}{4\pi}\right]=\varrho\fat{u}\cdot\fat{g}+n\Gamma-n^2\Lambda\\
 &\frac{\del}{\del t}\fat{B}+\nabla\times\left(\fat{B}\times\fat{u}\right)=0\\
 &\nabla\cdot\fat{B}=0\\
 &\nabla^2\Phi_\mathrm{gas}=4\pi G\varrho.\\
 \end{split}
\eeq
In the above set of equations, $\varrho,\fat{u}$ and $\fat{B}$ denote the gas mass density, the gas velocity and the magnetic field, respectively. The total energy density and pressure are given by 
\mbox{$E=\varrho\epsilon_\mathrm{int}+\varrho/2|\fat{u}|^2+1/(8\pi)|\fat{B}|^2$} and \mbox{$P_\mathrm{tot}=P_\mathrm{th}+1/(8\pi)|\fat{B}|^2$}, respectively. Furthermore, the gravitational acceleration 
\mbox{$\fat{g}=-\nabla\Phi_\mathrm{gas}+\fat{g}_\mathrm{sinks}$} consists of the acceleration from the gravitational potential of the gas and the acceleration by sink particles.
This set of equations is numerically solved by using the HLL5R Riemann solver to 
calculate the fluxes across cell boundaries \citep{Bouchut09,Waagan11} and a tree-solver to calculate the gravitational potential \citep[optimised for GPUs][]{Lukat16}.  For the (magneto-)hydrodynamics we apply 
periodic boundary conditions, while we use isolated ones for the gravity. This choice of mixed boundary conditions is purely of numerical convenience and does not affect any of our results as we only extract and use 
data far away from the boundaries, i.e., where the molecular cloud forms.\\
An overview of the main simulation parameters is given in Table~\ref{tabIC} and we refer the reader to \citet{Koertgen15} for a more detailled 
description of the initial setup.
\begin{table}
\centering
\caption{Summary of the main simulation parameters.}
\begin{tabular}{rl}
\hline
\hline
\fat{Edge length}	&256\,pc\\
\fat{Min. cell size} 	&0.015625\,pc\\
\fat{Flow length} &112\,pc\\
\fat{Flow radius}  &64\,pc\\
$\mathcal{M}_\mathrm{flow}$	&2\\
$\mathcal{M}_\mathrm{rms}$	&1\\
$\left(B_\mathrm{x},B_\mathrm{y},B_\mathrm{z}\right)_\mathrm{init}$  &(3,0,0)\,$\mu\mathrm{G}$\\
$T_\mathrm{init}$ 	&5000\,K\\
$n_\mathrm{init}$ 	&1\,$\mathrm{cm}^{-3}$\\
\hline
\hline
\end{tabular}
\label{tabIC}
\end{table}
\section{R\'{e}sum\'{e} of the general cloud evolution}\label{flows}
\begin{figure*}
 \begin{tabular}{ccc}
 $t=14.5\,\mathrm{Myr}$ &$t=20.5\,\mathrm{Myr}$ &$t=26.5\,\mathrm{Myr}$ \\
  \includegraphics[width=0.28\textwidth]{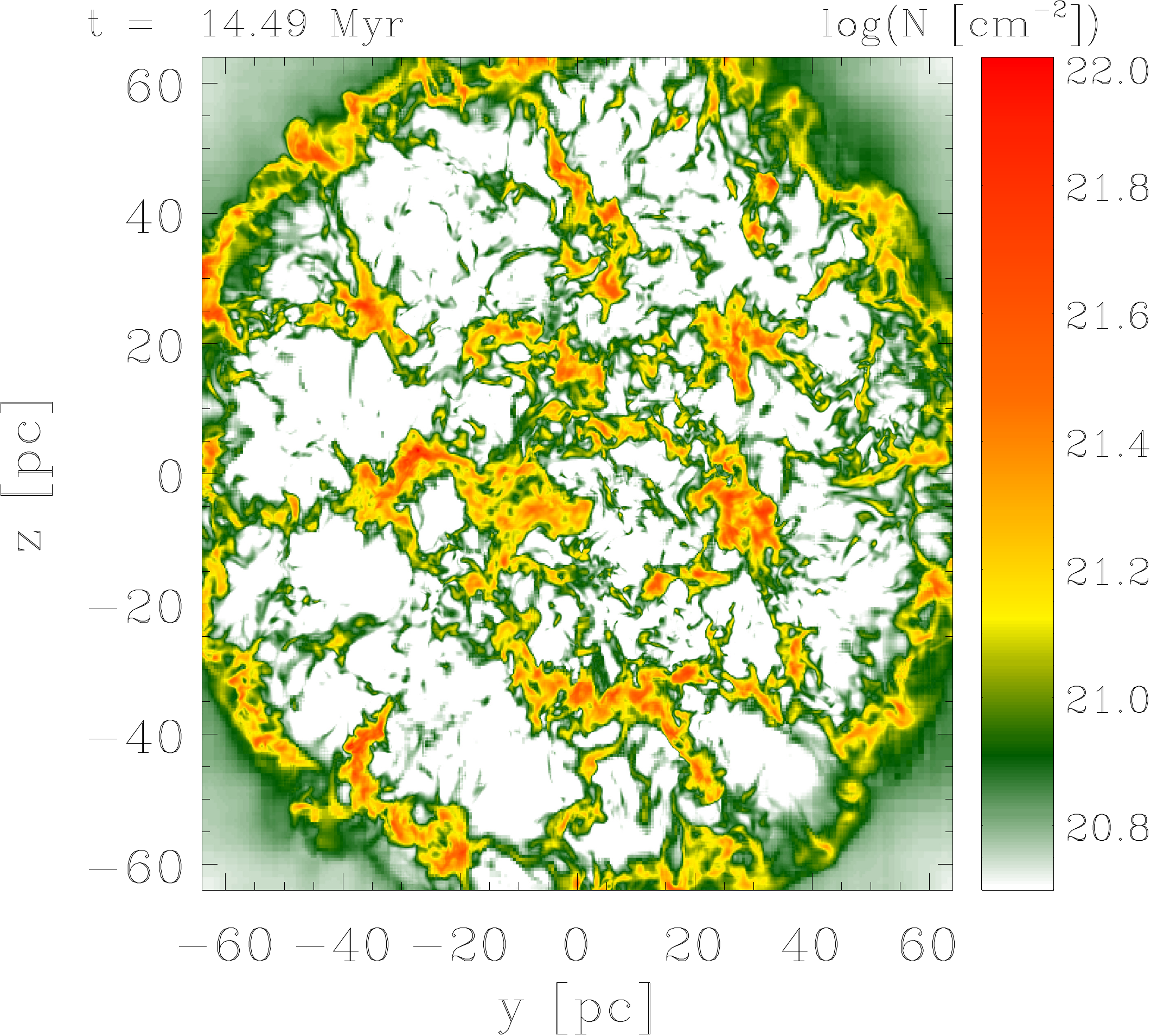} &\includegraphics[width=0.28\textwidth]{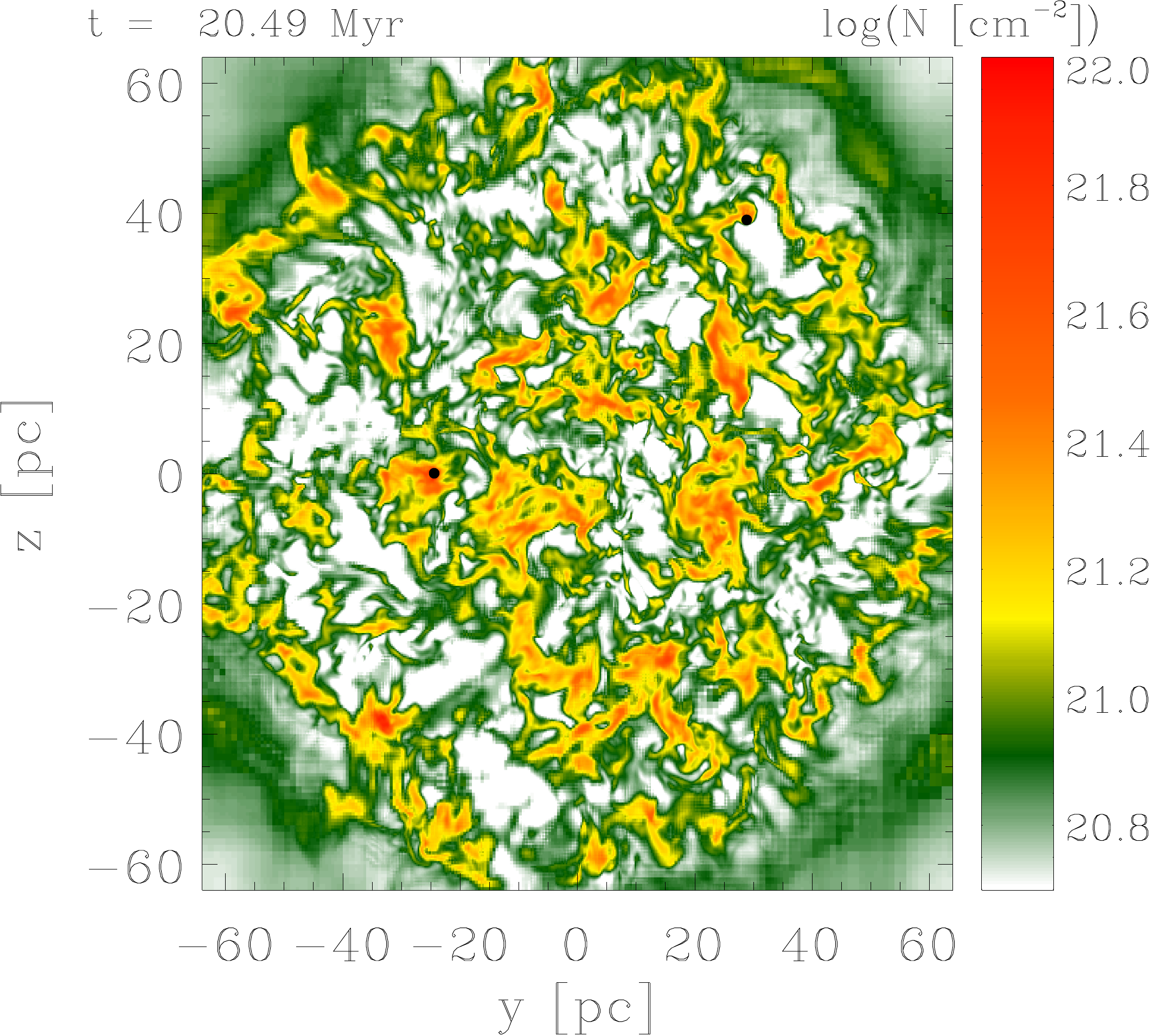}&\includegraphics[width=0.25\textwidth,angle=90]{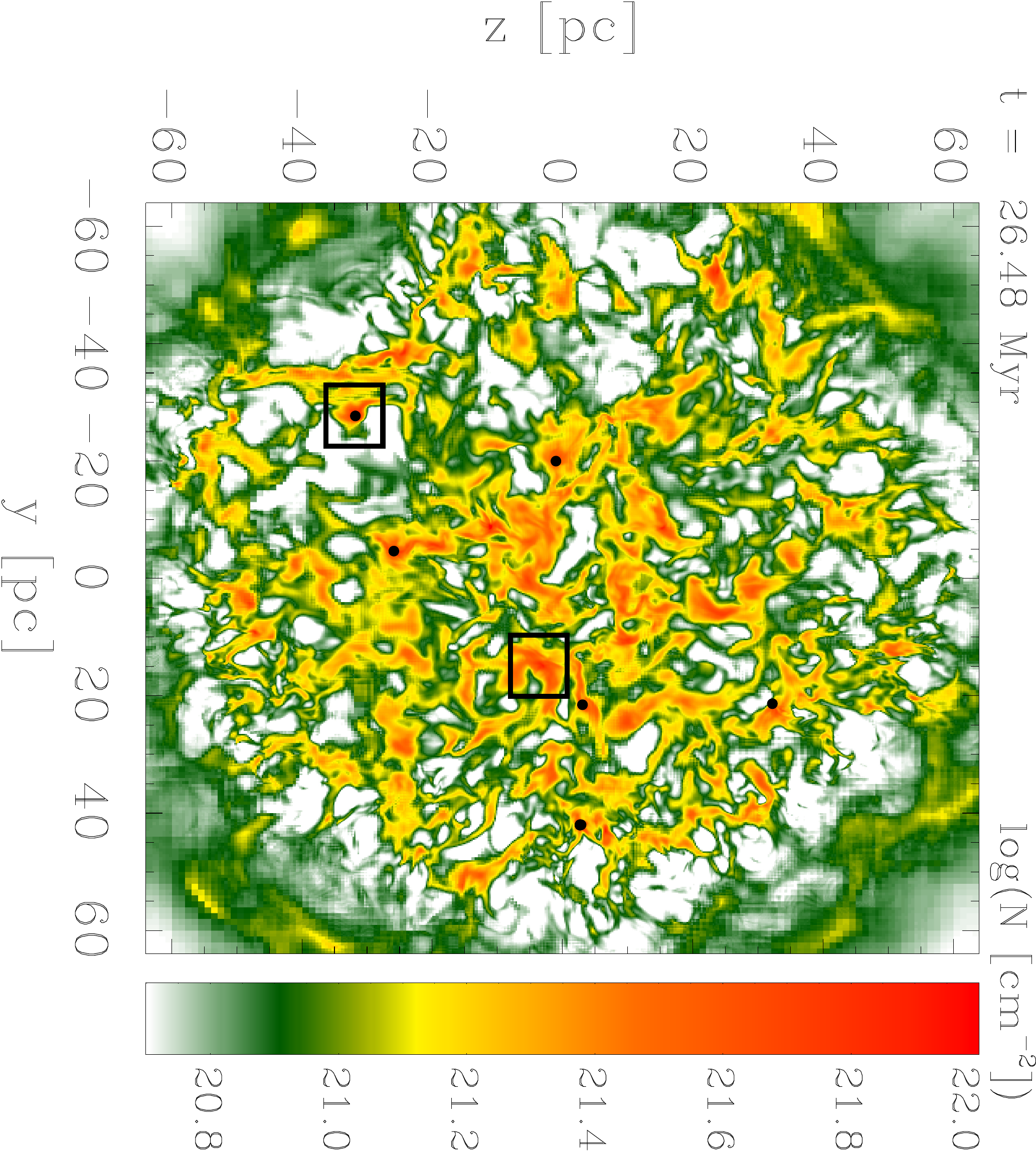}\\
  \includegraphics[width=0.28\textwidth]{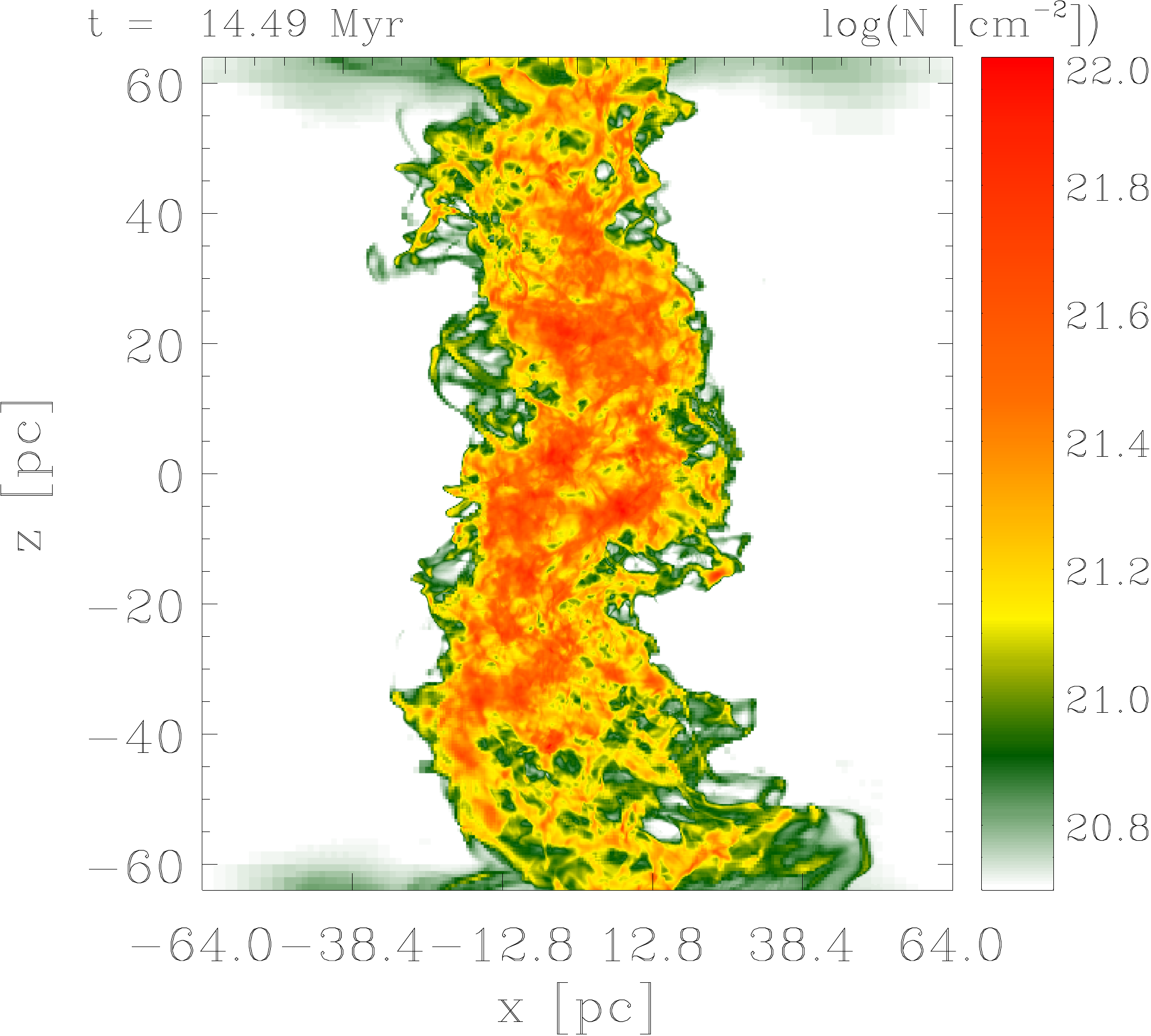}&\includegraphics[width=0.28\textwidth]{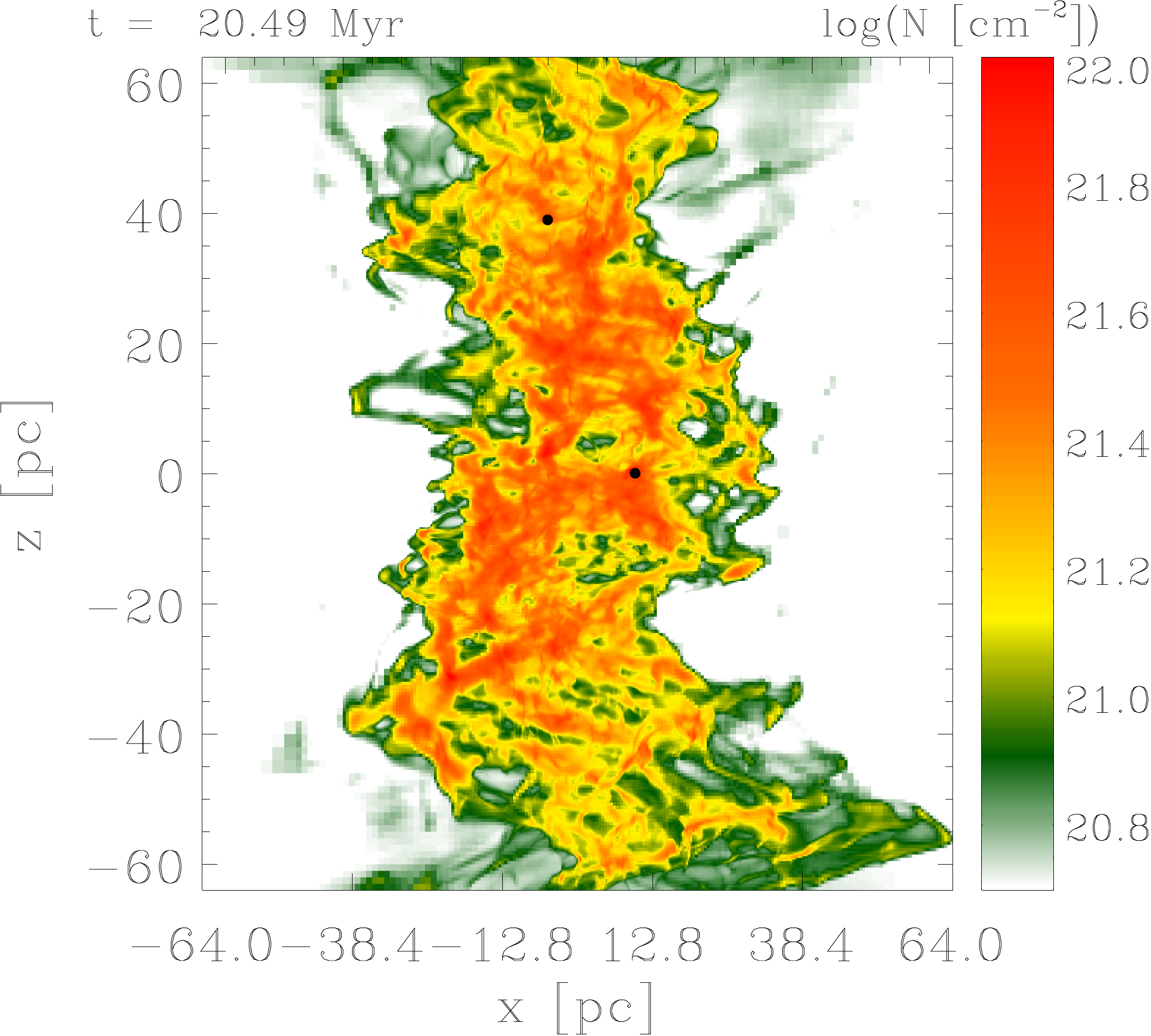}&\includegraphics[width=0.25\textwidth,angle=90]{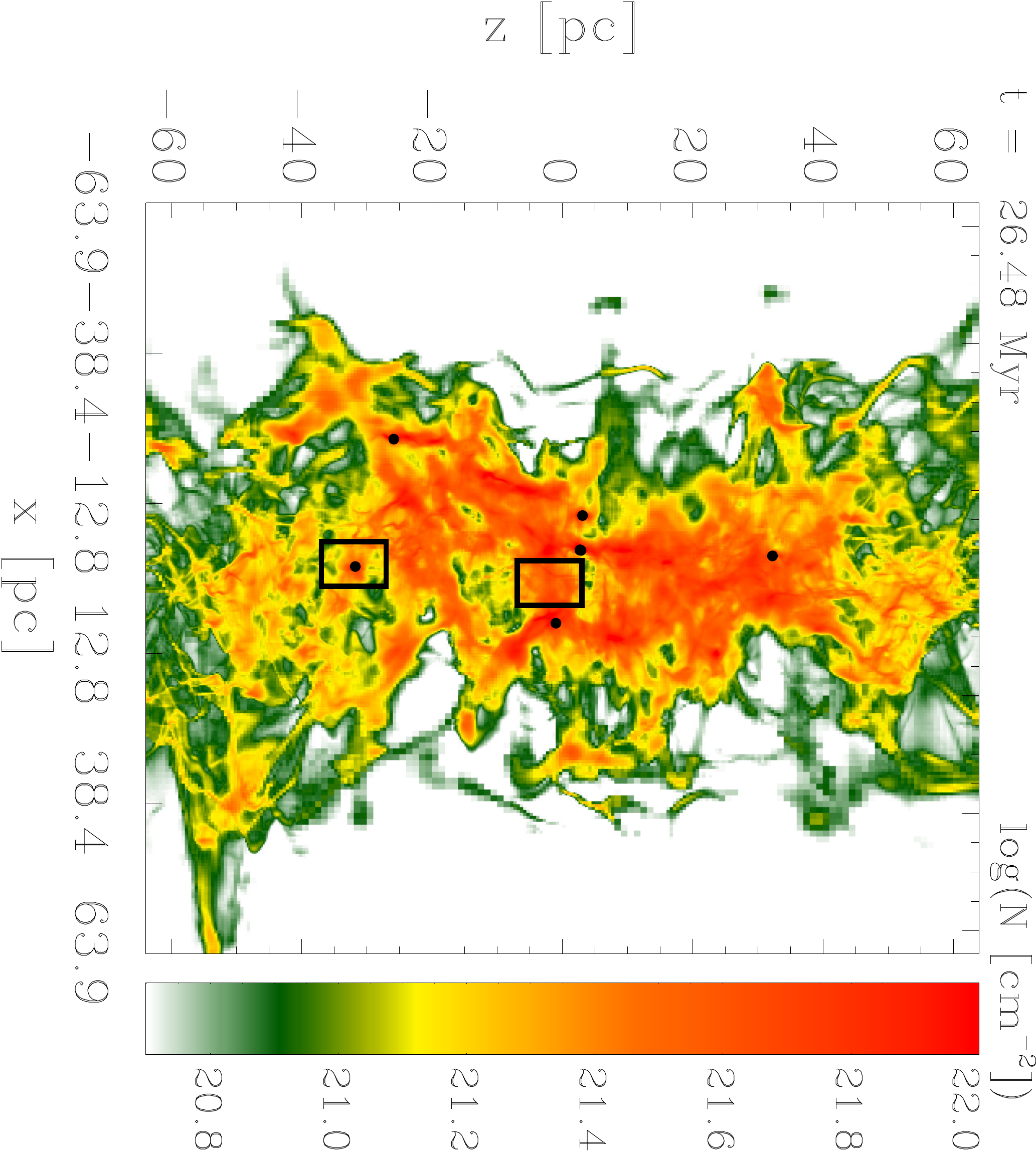}
 \end{tabular}
\caption{Column density maps of the cloud complex formed in between the WNM flows at different times (from left to right: 
14.5\,Myr, 20.5\,Myr, and 26.5\,Myr). The upper panels show 
the complex seen face-on, that is, along the flow direction. Bottom panels show an edge-on view. The black dots represent sink 
particles and the black boxes highlight the regions that will be analysed in more detail below. 
Integration length is $L=160\,\mathrm{pc}$ in all cases.}
\label{figGE}
\end{figure*}
In this section we briefly recap the evolution of the molecular clouds formed in the shocked slab between the converging WNM streams \citep[see e.g.][]{Banerjee09a,Koertgen16,Koertgen17}.\\
The time evolution of the cloud formed in the simulation, having an initial turbulent Mach number \mbox{$\mathcal{M}_\mathrm{RMS}=1.0$}, is shown in Fig.~\ref{figGE}. The (dynamical) time for the gas at the outer edges of the flows to reach the center is $t\sim10\myr$, so the column 
density maps present the more evolved stages of the cloud. The top row of Fig.~\ref{figGE} shows the cloud complex face-on, where the line of sight is along the initial flow direction. The bottom row depicts an edge-on 
view. Generally it is seen that the cloud is composed of filaments and dense clumps, which appear at the intersection of multiple filaments. At late times, the previously teneous regions between the filaments are 
seen to be denser since the cloud is collapsing globally \citep[see also][]{Vazquez17}. The turbulence generated by the inflows and by gravity shapes the morphology in the sense that the filamentary morphology prevails.\\
At $t=14.5\myr$ the cloud size is comparable to the size of the flows. 
The edge-on view reveals a bent morphology, which is the result of dynamical instabilities that are triggered by the turbulence within the flows. Over the course of the evolution, the cloud is observed to shrink in 
size, due to the aforementioned collapse on a global scale. As the free-fall time is much shorter for the denser parts on smaller scales \citep{Heitsch08c,Vazquez09,Hennebelle11,Federrath12}, hierarchical fragmentation has led to the 
formation of sink particles in the cloud interior. As more and more regions are collapsing the number of sink particles increases. At the last time shown, the number of sink particles has grown to 
$N_\mathrm{sink}=16$, where most of them are in close orbits around each other, because of further fragmentation of the parental core, so that only a handful are readily seen.\\
The outskirts of the cloud are observed to be highly irregular. Initially, dynamical instabilities distort the gas flows. At late times, the accretion flow onto the cloud is turbulent, which leads to further 
irregularity of the cloud boundaries \citep{Klessen10}. 
\subsection{The shape of the column density probability distribution function}
\begin{figure*}
	\centering
		\begin{tabular}{cc}
		\includegraphics[width=0.35\textwidth,angle=-90]{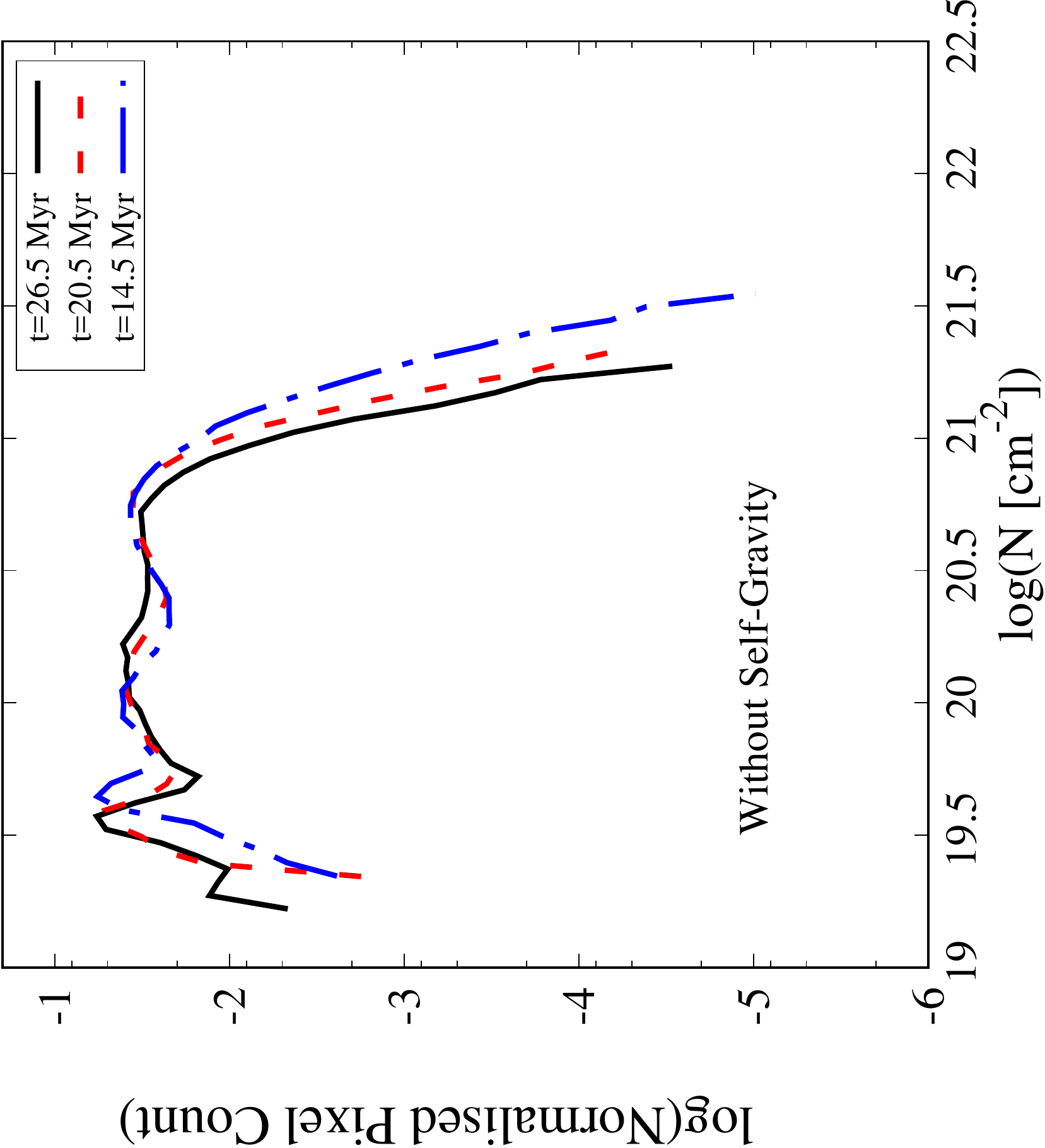}&\includegraphics[width=0.35\textwidth,angle=-90]{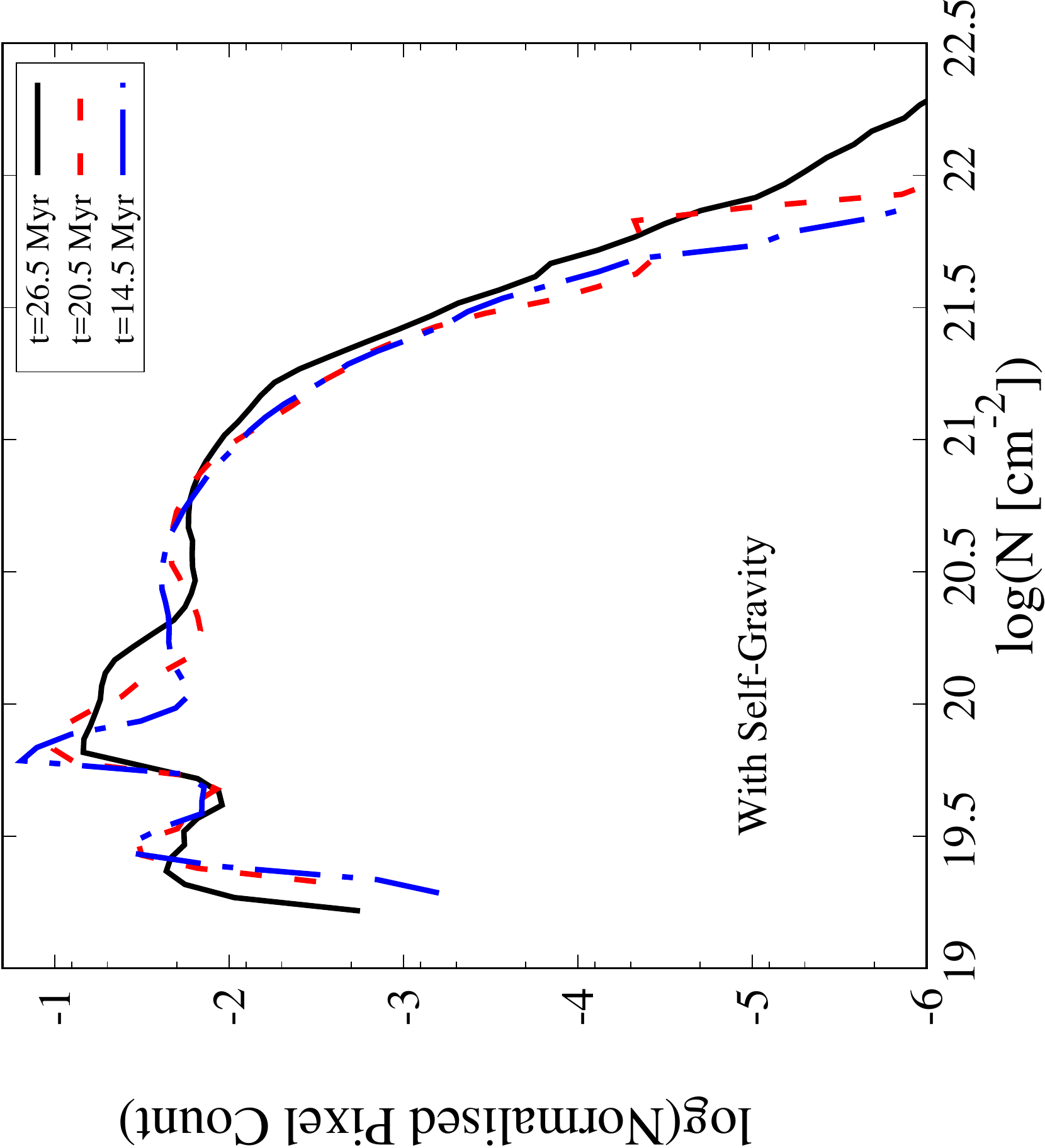}
		\end{tabular}
		\caption{Column density PDF at three times for the whole cloud complex in scenarios without (left) and with 
		self-gravity (right). Four differences are readily seen. At first, the peak of the PDF shifts to slightly higher 
		column densities for the case with self-gravity. Secondly, due to self-gravity, higher column densities are reached. 
		The shape of the PDF is composed of multiple peaks in the case \ita{without} self-gravity due to the greater 
		influence of thermal instability, whose signature appears to be washed out in the case with self-gravity. Last, in the 
		case with self-gravity, power-law tails emerge.}
		\label{figPDFcomp}
\end{figure*}

In Fig.~\ref{figPDFcomp} we show the column density PDF at different times and for simulations without and with 
self-gravity. We remind the reader that the dynamical time of the flows is $t\sim10\,\mathrm{Myr}$. The cloud evolution shown here is thus not influenced anymore by the initially coherent inflowing gas streams from the WNM.\\
The PDF of the non-gravitating gas is composed of multiple peaks at $\mathrm{log}(N/\mathrm{cm}^{-2})\sim19.6, 20.1$ and $20.7$, respectively. 
The leftmost peak at the lower column density is the surrounding medium. It shows a small 
width, because it is non-turbulent initially. The middle and right peaks show more evolution. At early times, \mbox{$t=14.5\,\mathrm{Myr}$}, 
the peaks appear to be well separated due to the action of thermal instability and the resulting differentiation into a 
multiphase medium \citep[see e.g.][]{Iwasaki14}. However, this double-peaked signature becomes weaker at later times \citep[see also][]{Vazquez00}. It is further seen that the PDF shifts 
towards lower column densities. Initially, gas is being compressed by the flows and starts to expand, once the compressing 
agent has vanished. The high-density part of the PDF appears to be not significantly different from a log-normal at any time. We point out that, in our case, the 
emergence of a power-law tail is thus solely due to gravity and not due to gas thermodynamics.\\
In contrast, the evolution of the PDF in the case with self-gravity is different. Firstly, with time, the PDF shifts towards higher 
column densities (in the high-density regime) due to the influence of self-gravity. Secondly, the emergence of a power-tail is already seen 
\citep[see also][who study the evolution of the N-PDF in a thermally bistable, gravitationally influenced medium]{Ballesteros11b}. Furthermore, the shape of the PDF in 
certain column density regimes appears to vary more over time (e.g. around $\mathrm{log}(N/\mathrm{cm}^{-2})\sim20$). For comparison, the non-self gravitating simulation shows (more or less) just a shift of the PDF, but no 
clear variation.

\section{Results}\label{results}
\subsection{Methodology of the cloud analysis}
We here present the results of our analysis on the column density PDF. In the upper right panel of Fig.~\ref{figGE} we highlight the studied regions and in Table~\ref{tabBox} we provide information on the details of the analysis. 
We focus on two regions: The first one is star-forming. One sink particle has already formed and the column density map indicates a 
strongly condensed region, forming a new sink particle in the future. The second region is a quiescent one with no 
sink particles and slightly lower, but still comparable column density to the first one. This region furthermore shows a centrally condensed column 
density structure. \\
For the analysis, we proceed as follows:
We first produce a column density map of the specific region. Next, we calculate column density contours with a contour spacing of 
\mbox{$\Delta\mathrm{log}(N/\mathrm{cm}^{-2})=0.01$} in square fields of view (FoV) of varying 
side-length centered 
around the center of the region and estimate the value of the \ita{last closed} contour. This is then highlighted in the column density map and in the 
N-PDF of the individual FoV. \\
\begin{table}
\centering
\caption{Center-position of the analysed regions and list of square boxes used for data analysis in each region.}
\begin{tabular}{rccc}
\hline
\hline
\fat{Region}	&X	&Y	&Z\\
			&(pc)&(pc)&(pc)\\
			\hline
\fat{Region 1}	&0&-27.7 &-32\\
\fat{Region 2}	&0	&14.75	&-3.75\\
\hline
\hline
\fat{Box edge length} $\left(\mathrm{pc}\right)$		&\fat{A}&\fat{B}&\fat{C}\\
\hline
\fat{Region 1} &10.5&5.25&3\\
\fat{Region 2} &16&8&4\\
\hline
\hline
\end{tabular}
\label{tabBox}
\end{table}

\subsection{The completeness of the N-PDF}
\begin{figure*}
 \begin{tabular}{ccc}
  \includegraphics[height=0.225\textheight]{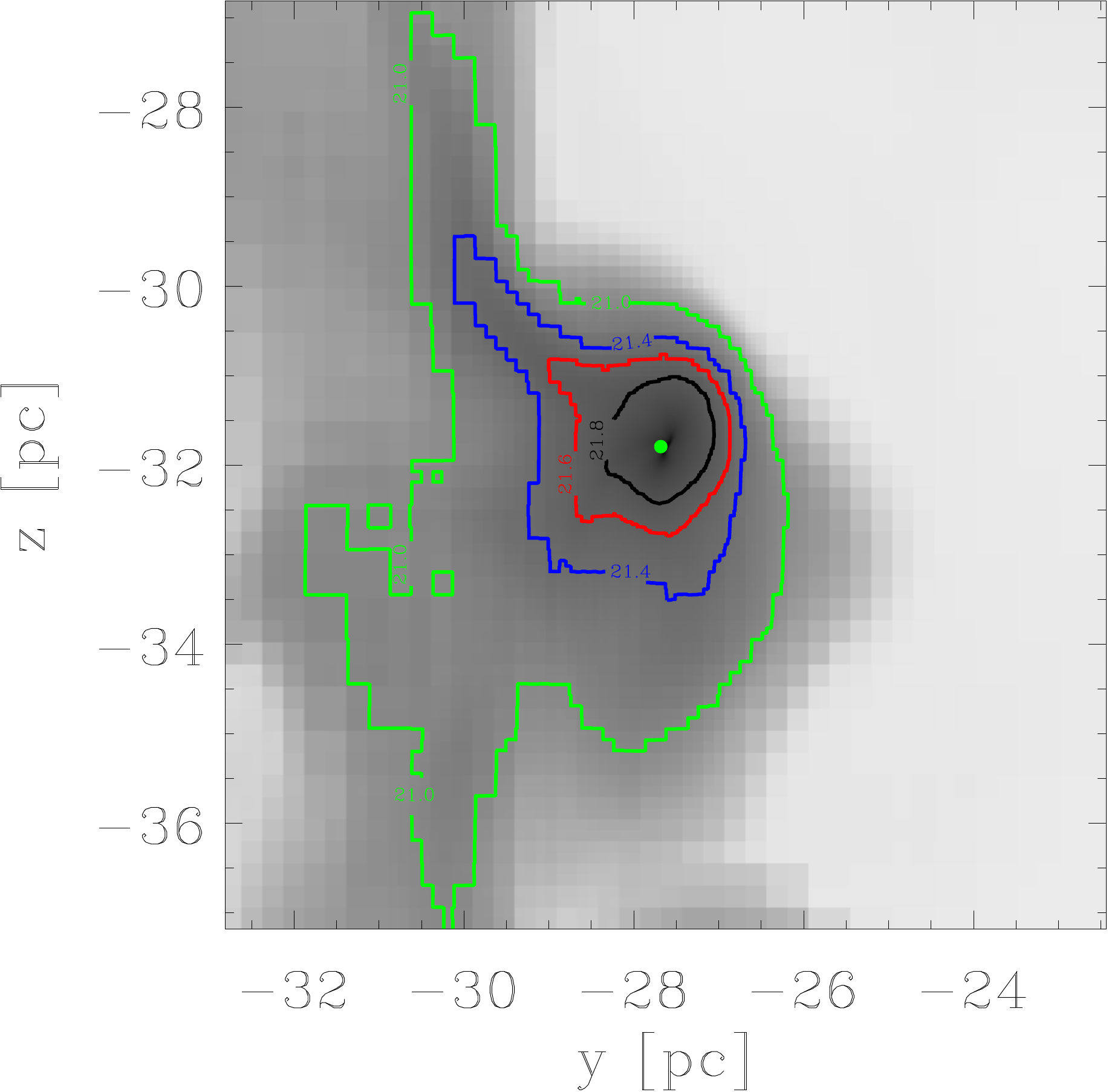}&\includegraphics[height=0.225\textheight]{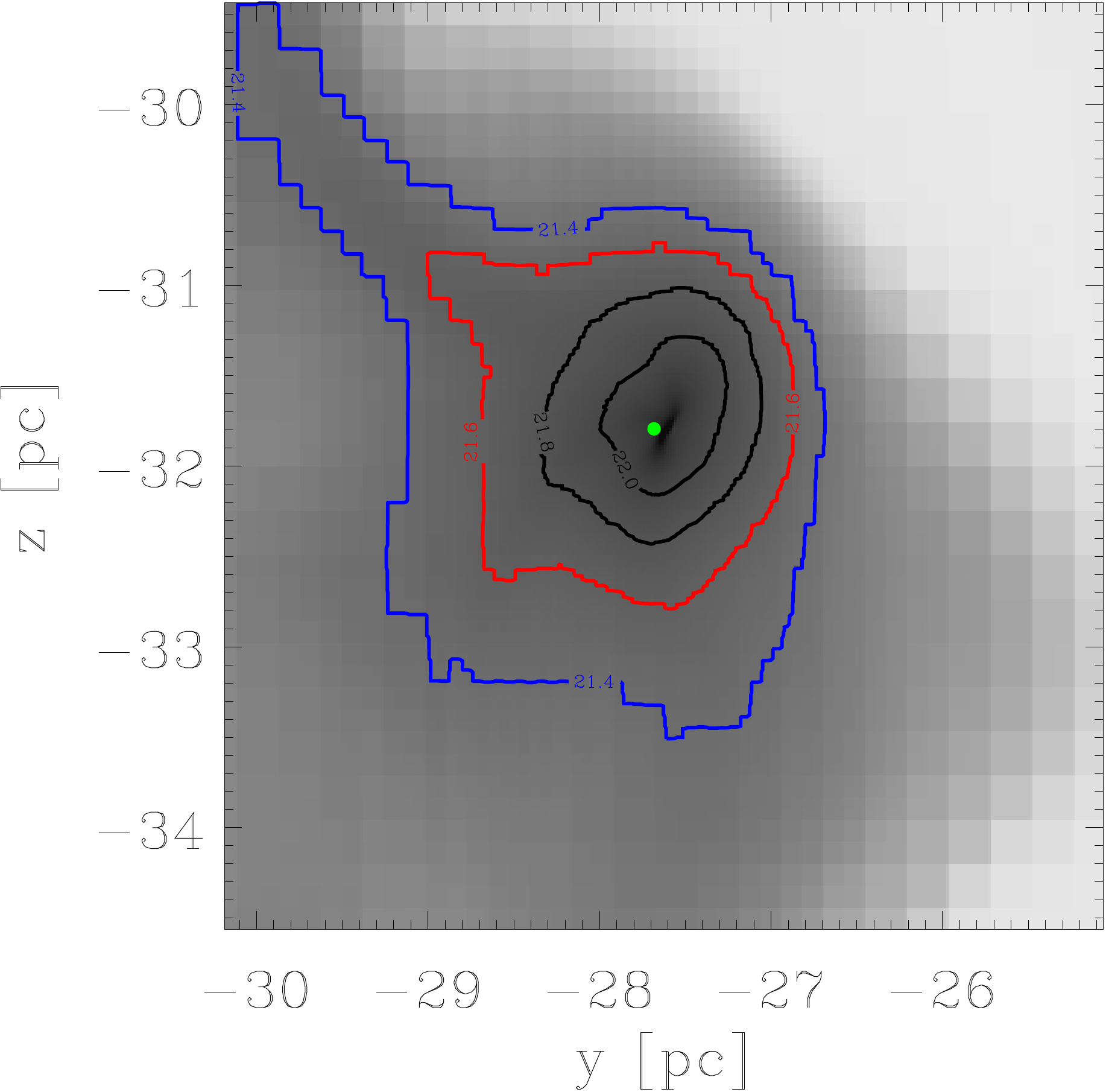}
  &\includegraphics[height=0.225\textheight]{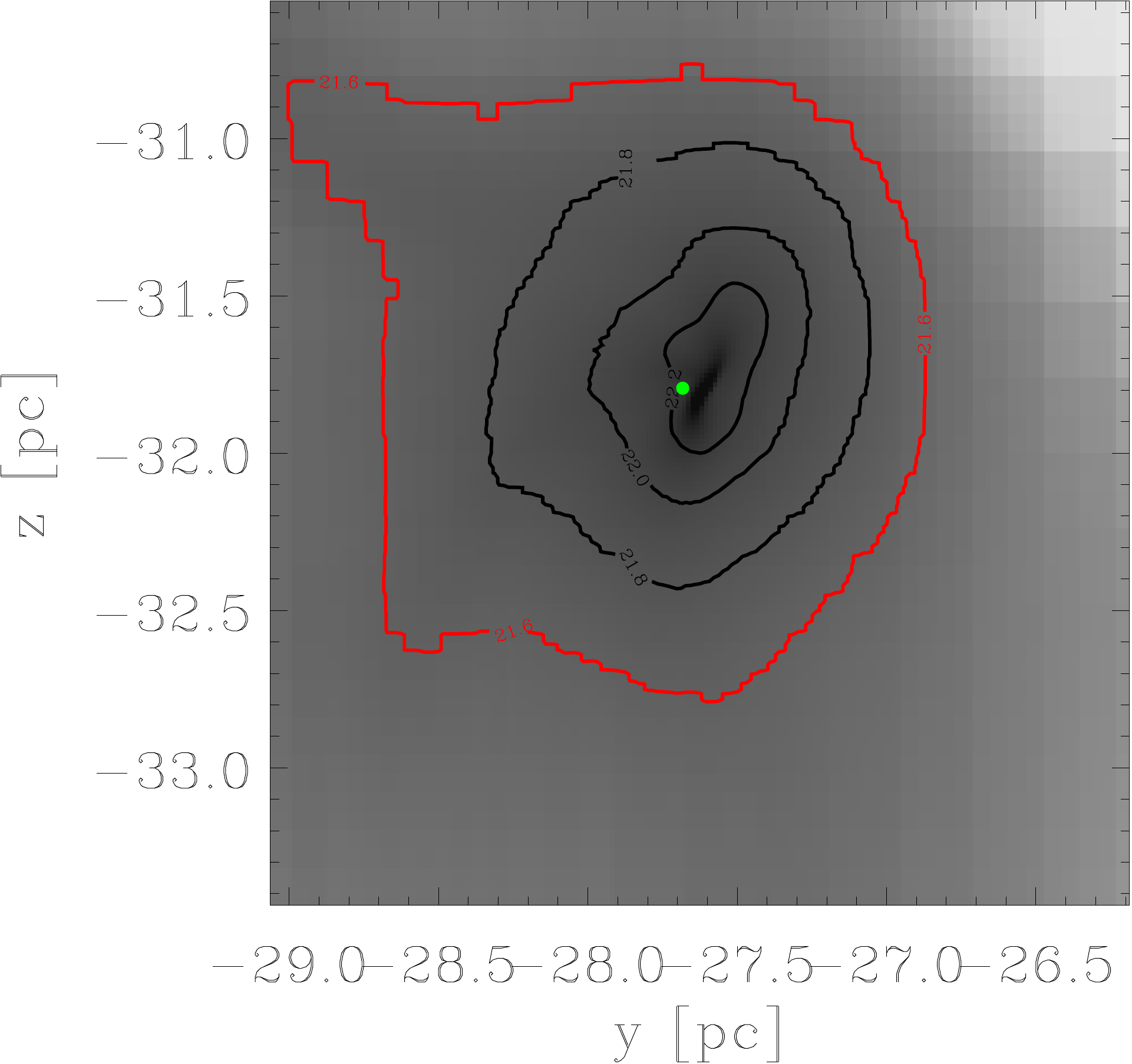}\\
  \includegraphics[width=0.33\textwidth,angle=-90]{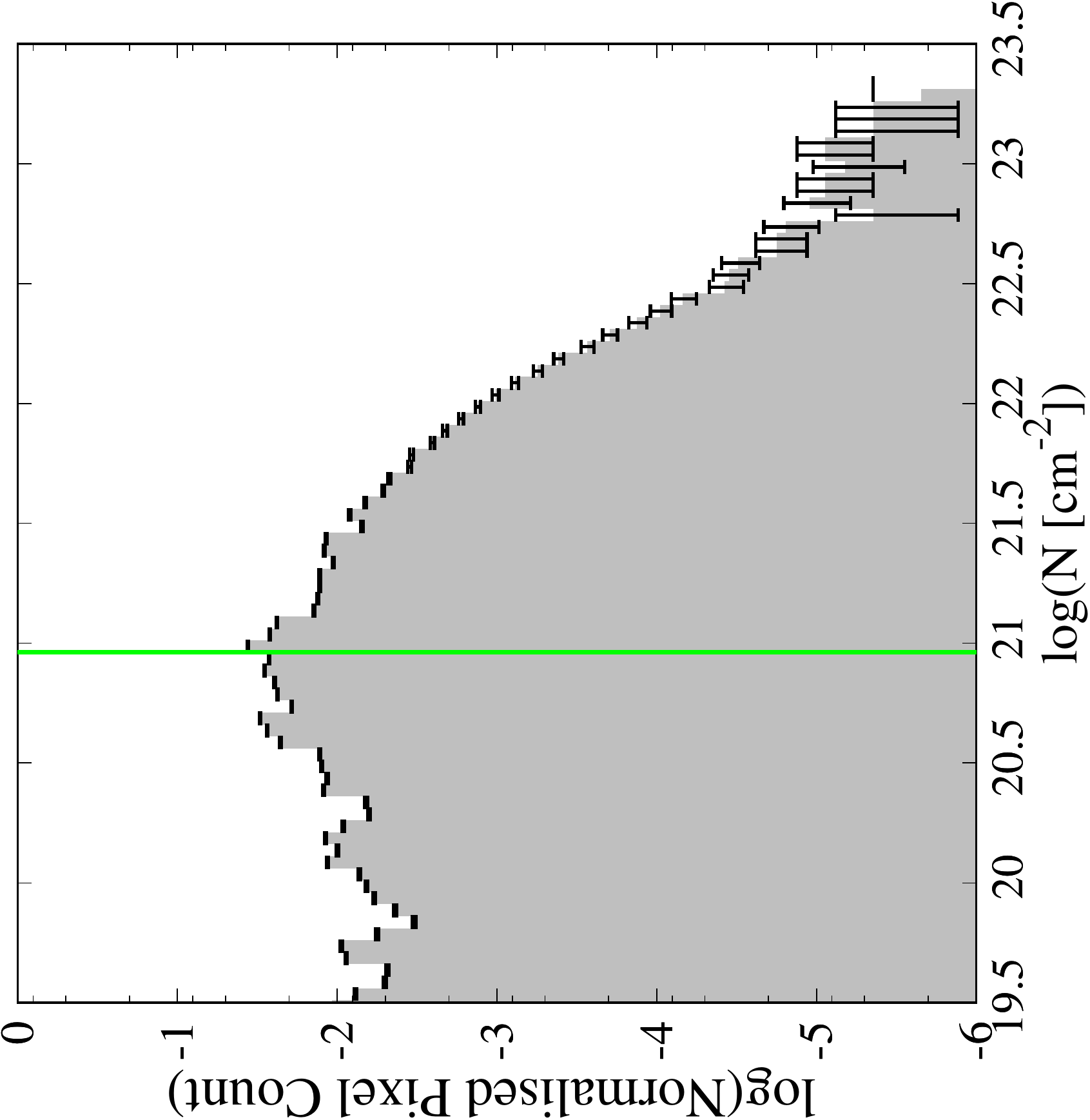}
  &\includegraphics[width=0.33\textwidth,angle=-90]{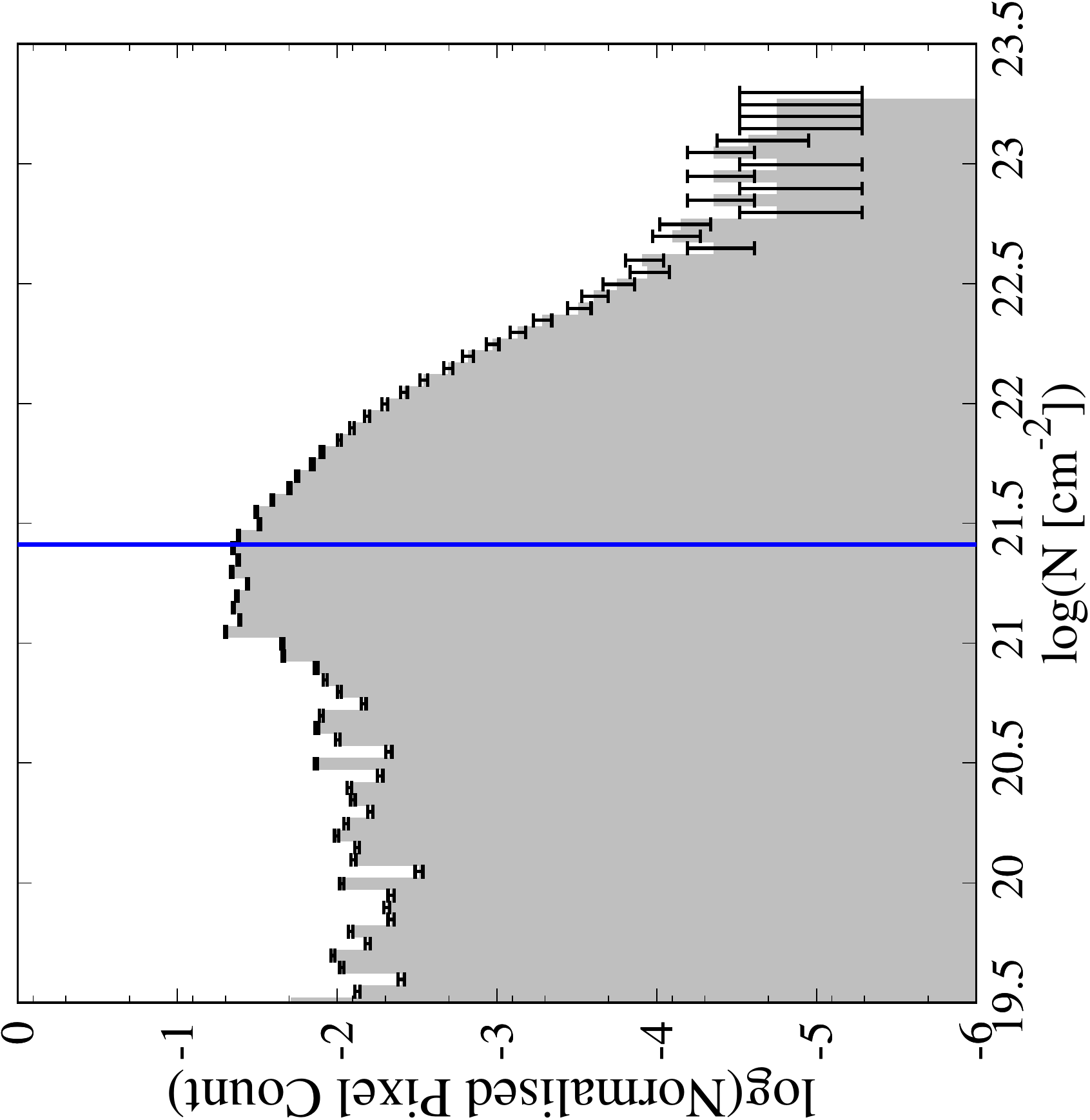}
  &\includegraphics[width=0.33\textwidth,angle=-90]{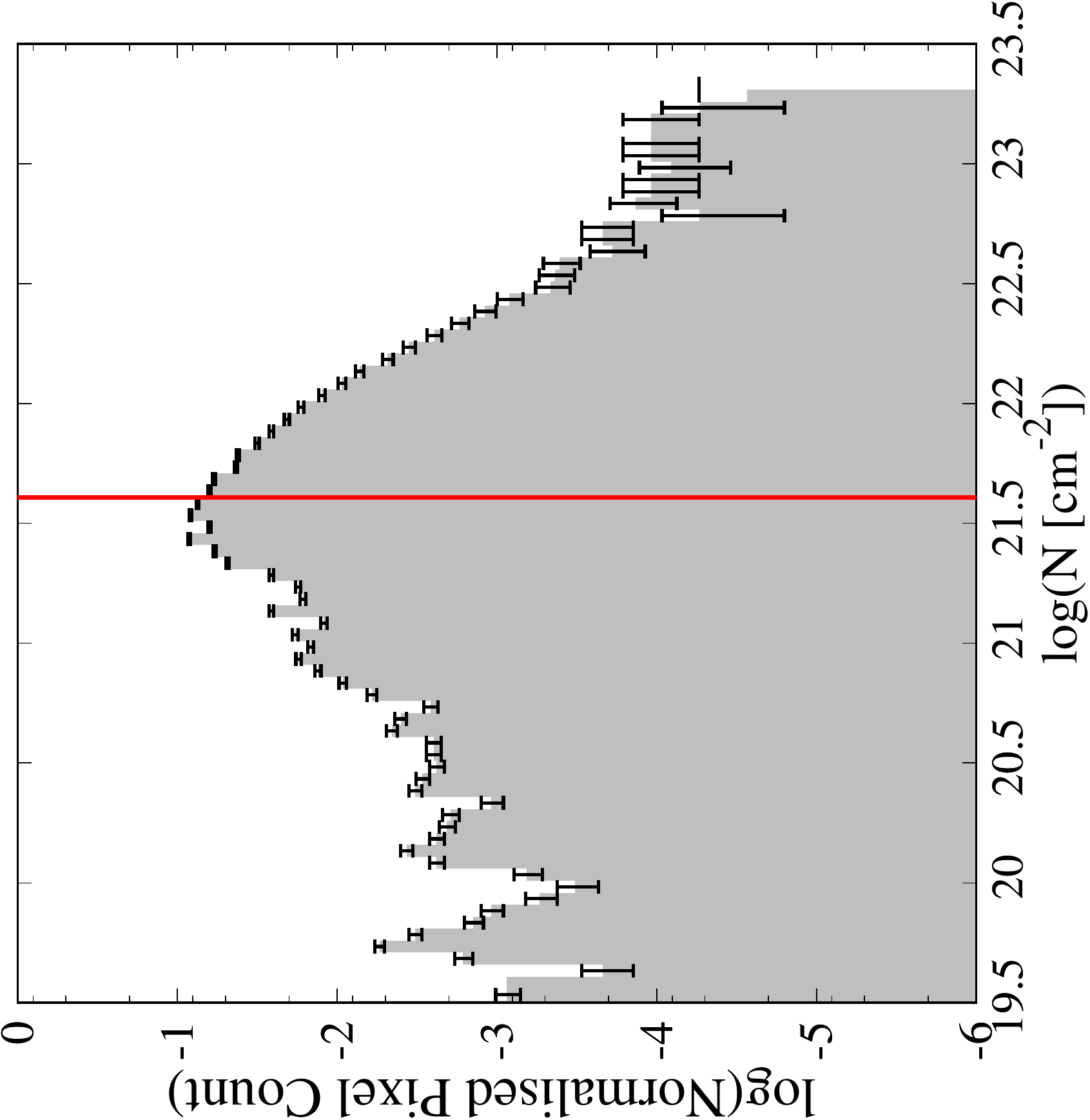}
 \end{tabular}
 \caption{\ita{Top:} Column density maps of a region undergoing gravitational collapse with decreasing field of view from 
$10\times10\,\mathrm{pc}^2$\,(left), over $5\times5\,\mathrm{pc}^2$\,(middle) to  $3\times3\,\mathrm{pc}^2$\,(right). The data range is 
 \mbox{$19\,(\mathrm{white})<\mathrm{log(N\,\left[\mathrm{cm}^{-2}\right])}<23\,(\mathrm{black})$}. Overlaid are contour lines of 
 the column density. The green, blue and red contour lines indicate the \ita{last closed contour} for the largest, intermediate and smallest 
 field of view, respectively. Black lines denote closed contours with contour spacing of $\Delta\mathrm{log}(N)=0.2$. \ita{Bottom:} Corresponding column density PDF (grey) with overlaid value of the last closed contour, which highlights the completeness limit. The black error bars denote the Poisson errors of each bin, calculated as 
 $\sqrt{N_\mathrm{bin}}$ with $N_\mathrm{bin}$ being the pixel count of each bin. It is clearly seen that, for a larger field of view, 
 more and more of the PDF, which deviates from a power-law falls within the completeness limit. Note the gradual shift of the value of the last closed contour towards higher column densities for decreasing field of view.}
 \label{figPDF}
\end{figure*}

In Fig.~\ref{figPDF} we show column density maps of the star-forming region 1 with 
overlaid column density contours. Some information can be found in \mbox{Tab.~\ref{tabBox}}. 
\begin{figure*}
 \begin{tabular}{ccc}
 \includegraphics[width=0.33\textwidth]{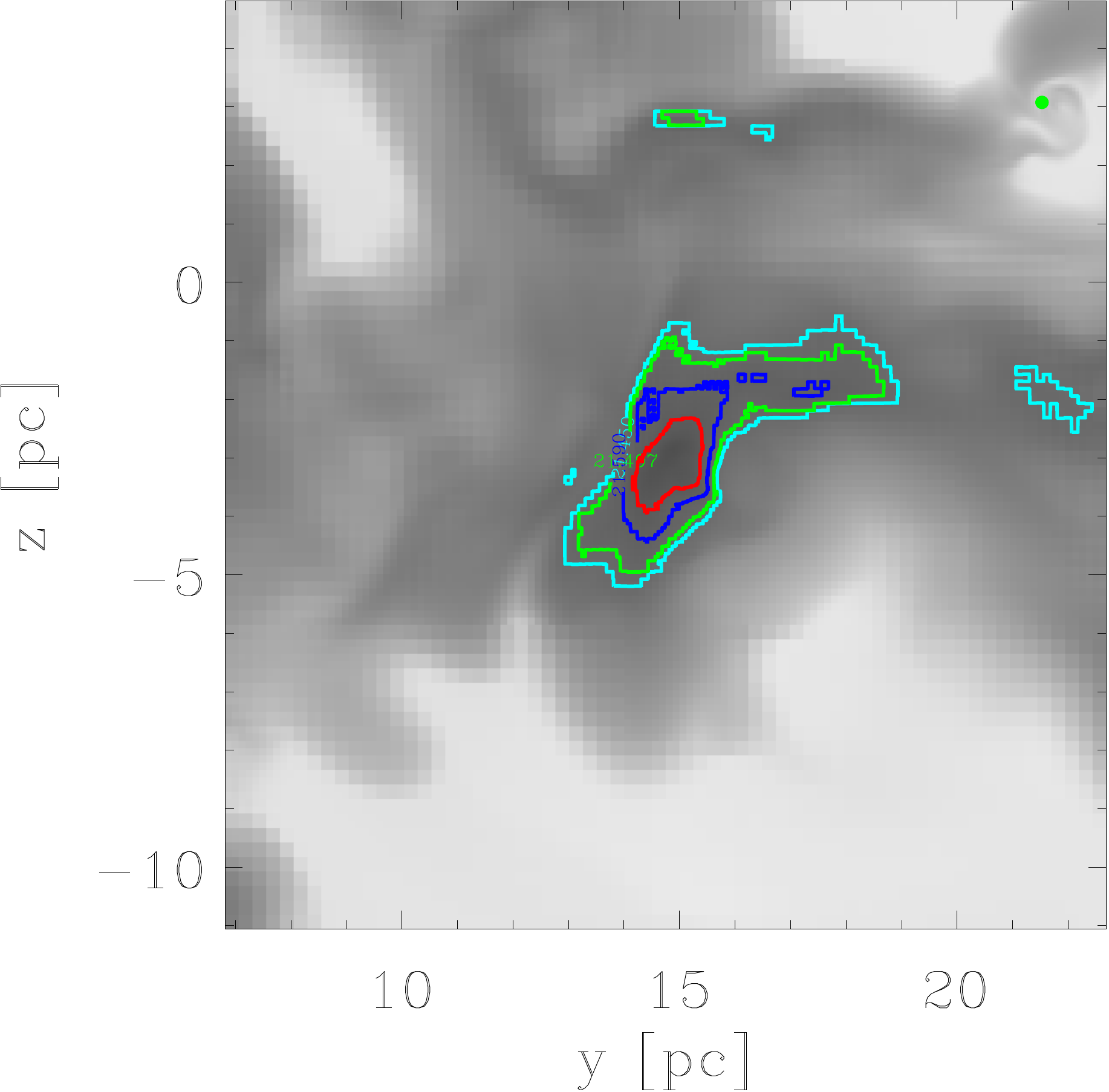}
  &\includegraphics[width=0.33\textwidth]{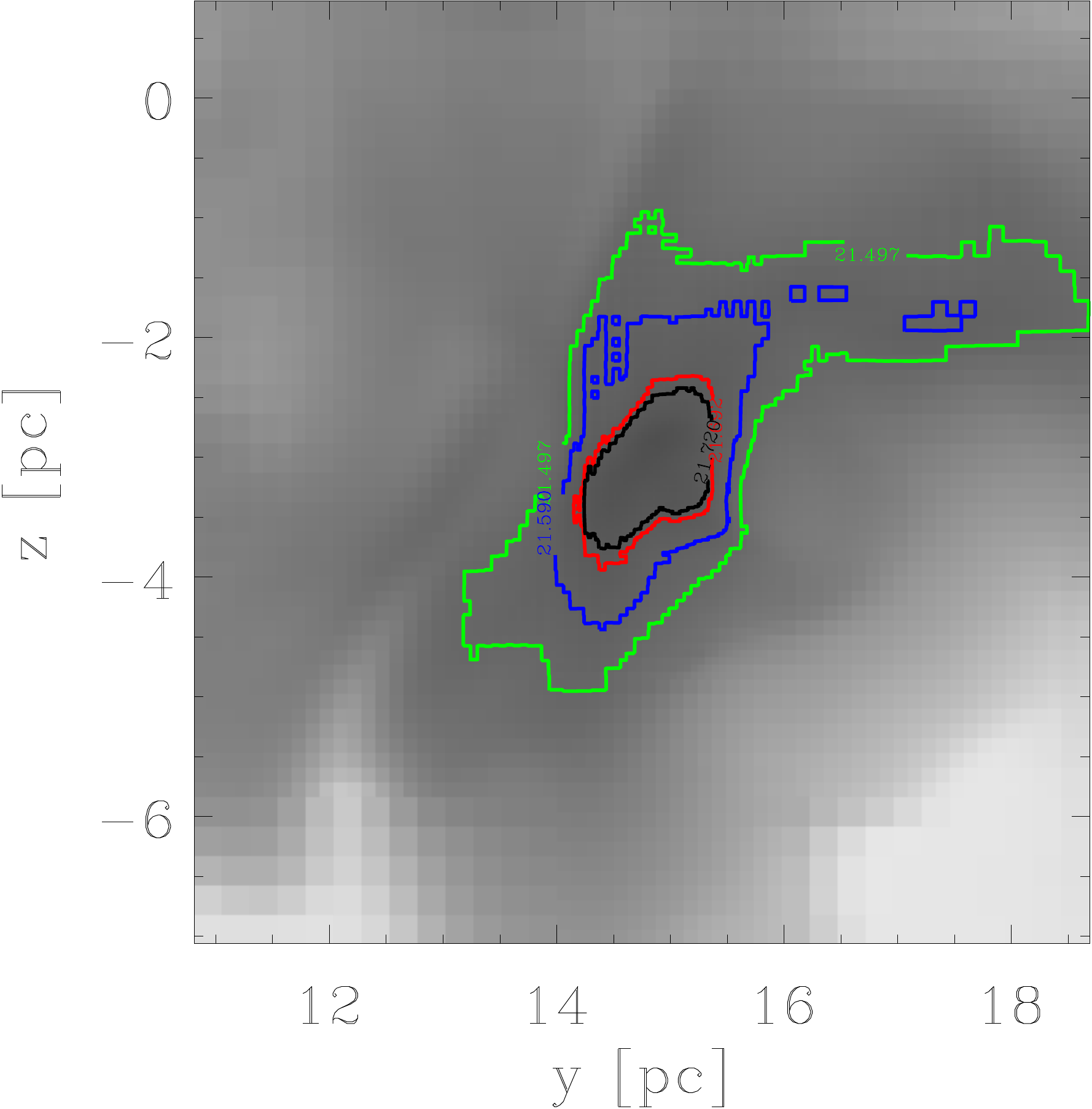}&\includegraphics[width=0.33\textwidth]{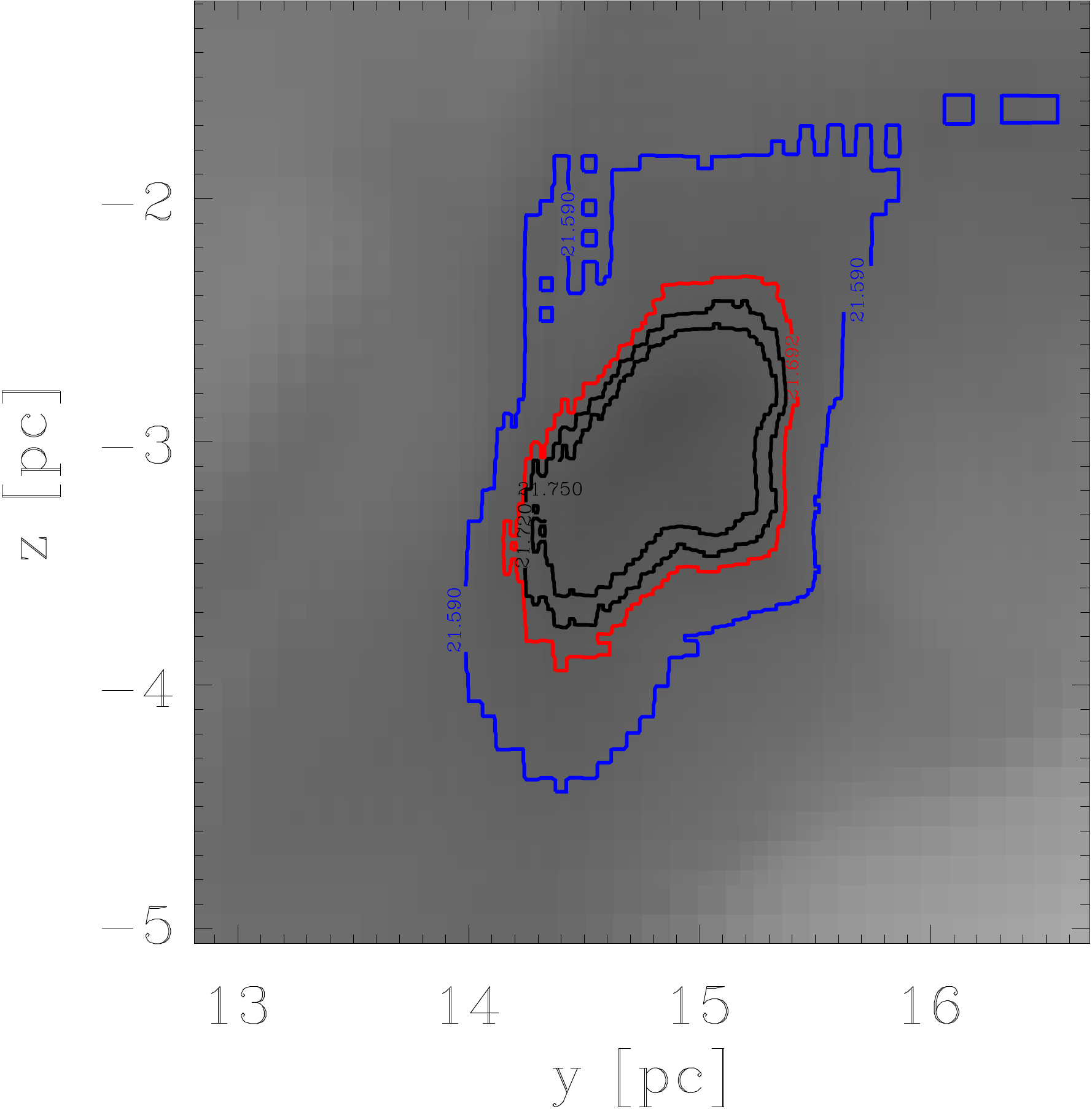}\\
  \includegraphics[width=0.33\textwidth,angle=-90]{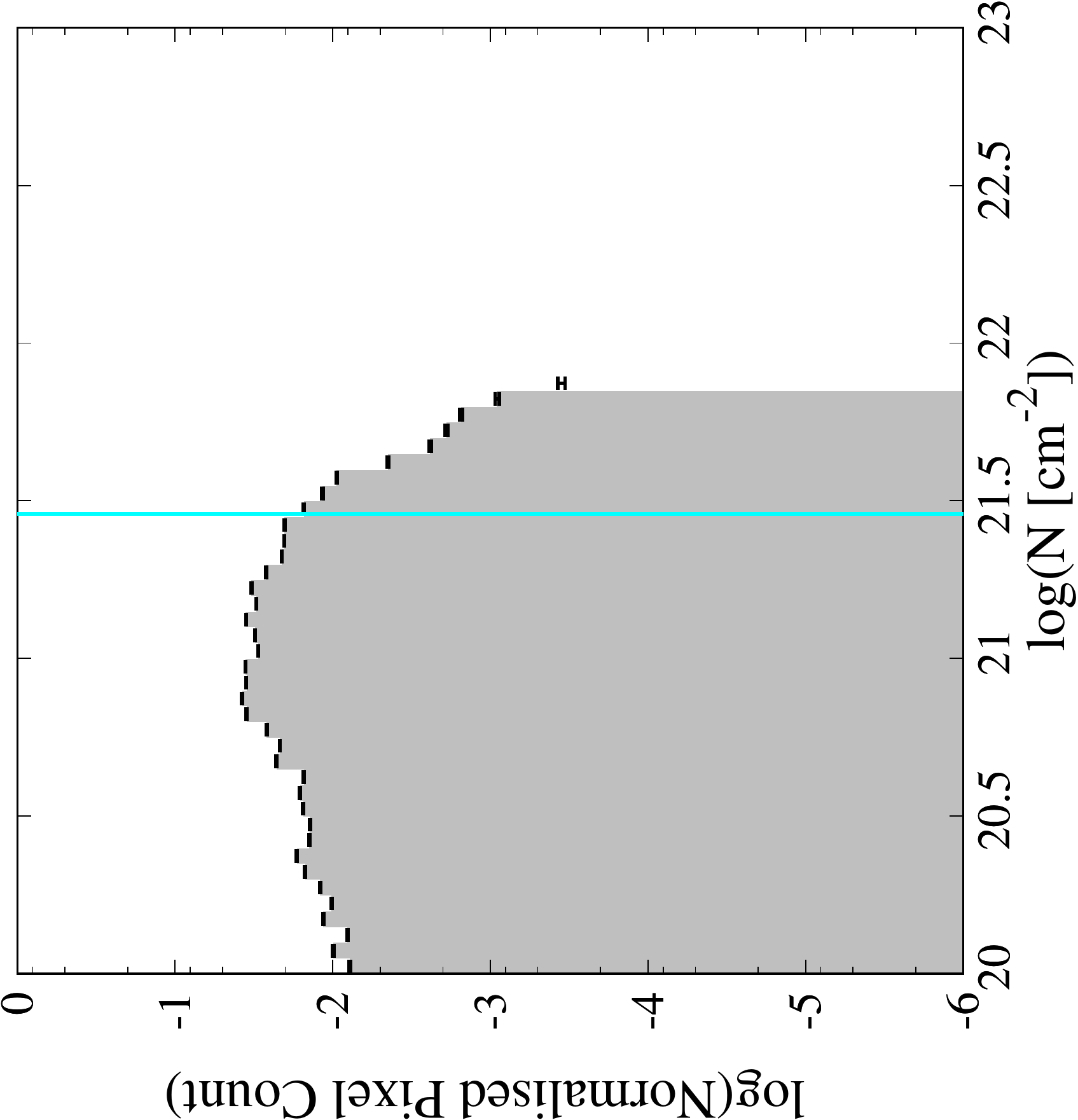}
 &\includegraphics[width=0.33\textwidth,angle=-90]{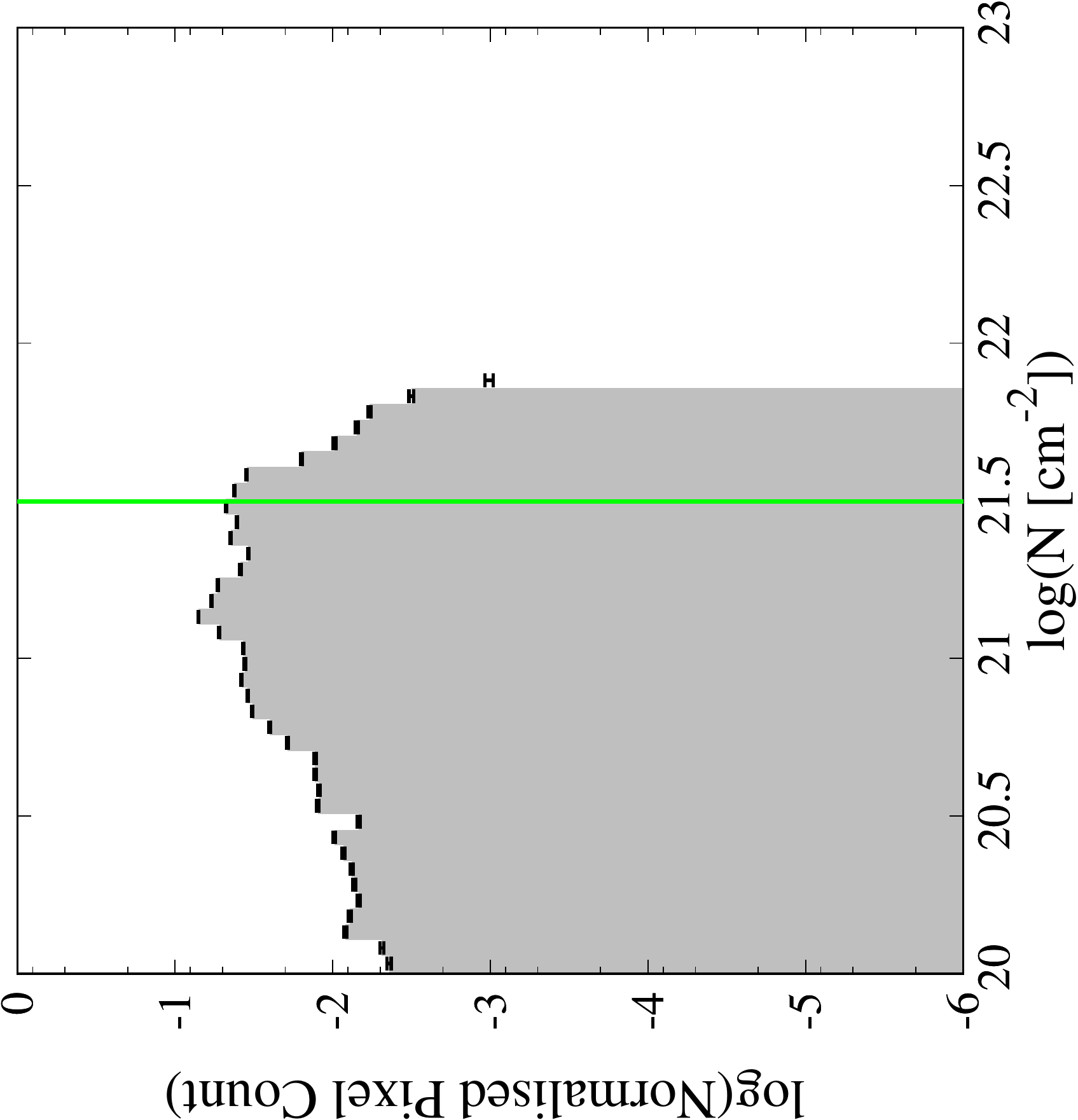}
  &\includegraphics[width=0.33\textwidth,angle=-90]{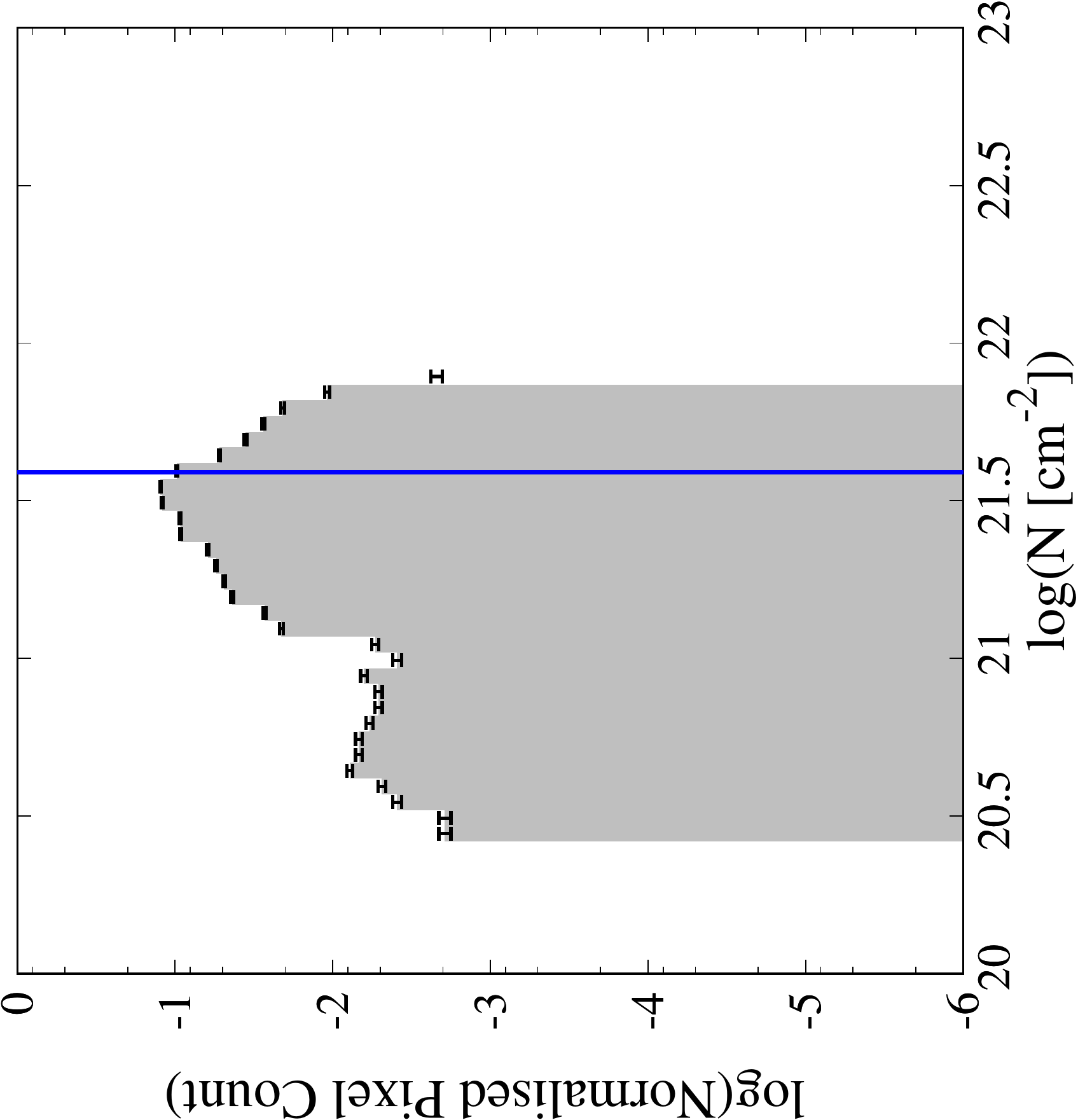}
 \end{tabular}
 \caption{Same as Fig.~\ref{figPDF} for a centrally condensed, but non-star-forming region. Please note that the contour spacing for the 
 black contours has changed as the column density profile is rather flat near the center. Note that, in the left panel with a field of view of 
 $16\times16\,\mathrm{pc}^2$, turbulent motions in the cloud create regions of 
 enhanced density at various positions within the field of view. This will lead to a last closed contour at higher column densities, if a high density region is located 
 near the boundaries of the field of view.}
 \label{figPDF2}
\end{figure*}
The FoV decreases from left to right, covering areas of $10\times10\,\mathrm{pc}^2$, $5\times5\,\mathrm{pc}^2$ and $\sim3\times3\,\mathrm{pc}^2$. This 
region is composed of a centrally condensed part near the already formed sink particle, which is embedded in a larger-scale filamentary 
structure (see also Fig.~\ref{figGE}). The coloured contour lines depict the last closed contour and we additionally show them within the 
next larger FoV. In the bottom row of the same figure, we show the column density distribution of the individual FoV. The value 
of the last closed contour is indicated by the coloured vertical line. It is evident that the value of the last closed contour shifts towards higher 
column densities, being at $\mathrm{log}(N/\mathrm{cm}^{-2})=20.9$ for the largest and at $\mathrm{log}(N/\mathrm{cm}^{-2})=21.6$ for the smallest FoV. Furthermore, 
it is evident that the distribution \ita{above} this value changes significantly. For the smallest FoV, it resembles a power-law tail, as this 
area is dominated by gravity. A small change in the slope of the power-tail is also observed for $\Delta\mathrm{log}(N/\mathrm{cm}^{-2})\sim22.7$, probably being 
indicative of a rotationally supported structure on the smallest scales \citep[e.g.][]{Kritsuk11a}. Increasing the field of view results in a slight shift of the 
value of the last closed contour to smaller column densities. Also for this FoV, the distribution above the last closed contour more closely resembles 
that of a power-law distribution, though the last closed contour is now seen to be closer to the turnover of the distribution. Increasing the 
field of view even further results in a larger shift of the last closed contour towards even smaller column densities. It is now evident that the 
\ita{resolved} distribution above the closed contour \ita{does not resemble a pure power-law distribution} anymore. This distribution also 
includes part of the log-normal turnover, though we emphasise that it does not completely fall within the completeness limit.\\
For comparison, we show in Fig.~\ref{figPDF2} the column density maps and corresponding distributions for a non-star-forming region. This 
region appears to be also centrally condensed, with the column density profile being rather flat near the centre. The FoV is 
$16\times16\,\mathrm{pc}^2$, $8\times8\,\mathrm{pc}^2$ and $4\times4\,\mathrm{pc}^2$. We point out that, in this region, it is not clear how to interpret the shape of the column density distribution at larger column densities. However, it is still evident that the 
value of the last closed contour shifts towards lower column densities with increasing FoV, which shows that the above stated argument 
also holds for other regions within the formed cloud complex.

\subsubsection{The last closed contour in a turbulent environment}
\begin{figure}
	\centering
		\includegraphics[width=0.4\textwidth,angle=-90]{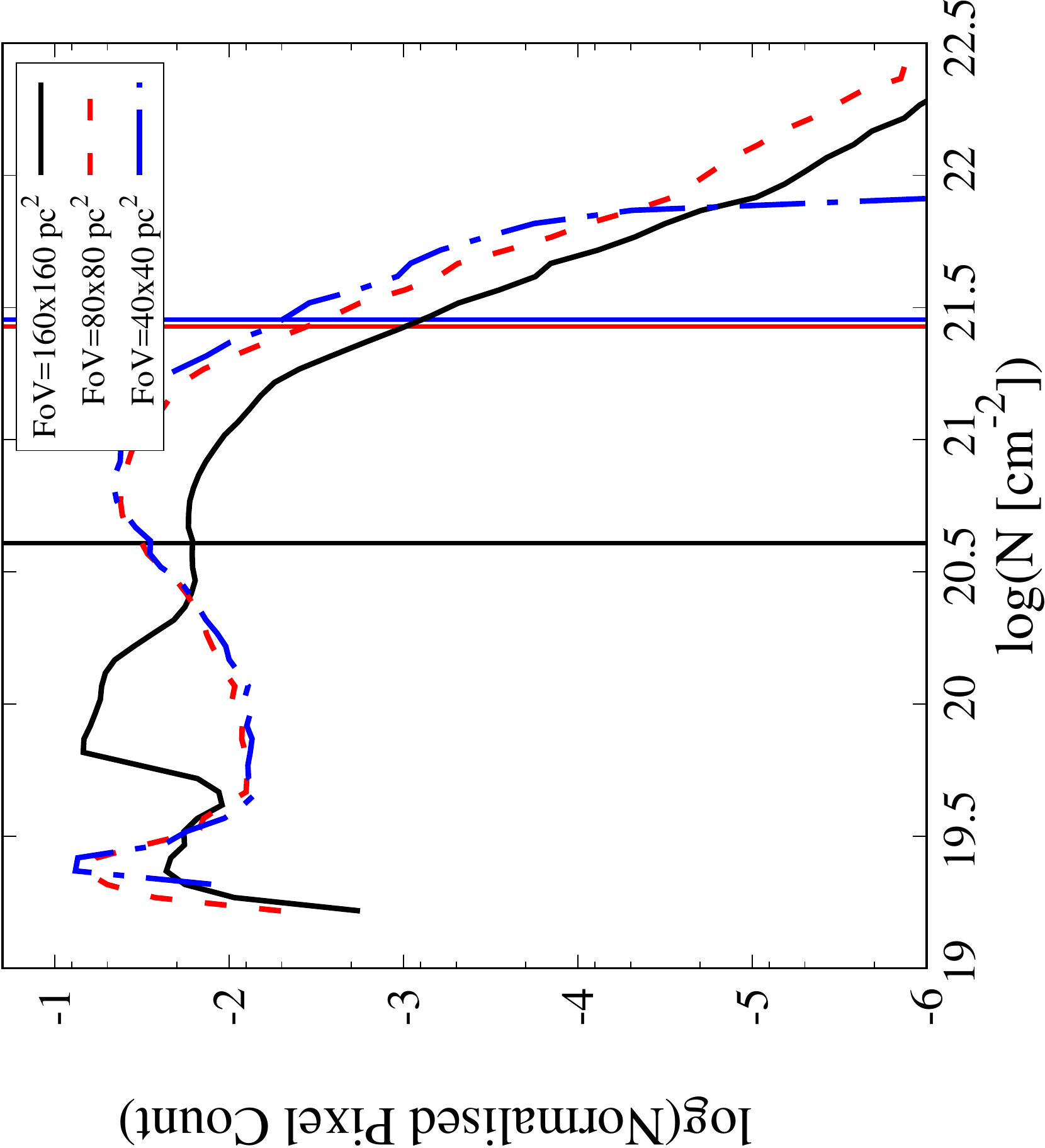}
		\caption{Distribution of column densities for three fields of view centered on the center of the simulation domain. Only the largest 
		field of view contains the whole cloud complex. The smallest field of view also excludes certain star-forming regions. Similar to the 
		individual regions within the complex, the turnover in the distribution is only captured for a sufficiently large FoV.}
		\label{figPDFwclcc}
\end{figure}

So far we have studied the last closed contour for relatively centrally condensed regions. The left panel in Fig.~\ref{figPDF2} highlights the effects of a turbulent 
environment, in which various overdensities appear within the FoV. If these regions are of similar column density or a high column density 
region resides near the boundaries of the field of view, this will lead to a value of the last closed contour that is at column densities well 
within the power-law tail.  Instead, one has to consider that in a fully turbulent medium, there will essentially be no closed contour anymore for a sufficiently low column-density threshold. This is just natural for a turbulent medium, so the current way of using closed contours to define a column-density threshold basically excludes the log-normal peak by construction. This applies in particular for regions, where only a single last closed contour is accepted within the FoV. Such single closed contours can only be the result of gravitational collapse and hence those will only capture the very highest-density part of the PDF, where gravitational collapse has turned the high-density wing of the original log-normal PDF into a power-law tail. \\
 Fig.~\ref{figPDFwclcc} shows the column density 
distribution for a varying field of view for the whole cloud complex that has been formed in the compression layer of the two converging, 
turbulent WNM flows. Highlighted by vertical lines are the values of the last closed contour for the given field of view. It is obvious that, in the 
case of a turbulent, and thus patchy, environment, the last closed contour is well within the power-law regime of the distribution. Only the 
largest field of view, which captures the entire initial radial extent of the WNM flows and some diffuse gas, is able to also resolve the 
turnover in the distribution. We, however, caution here that this is based on our rather idealised simulation setup and things will look differently in the real ISM.

\section{Conclusions}\label{conclusions}
We have presented results from molecular cloud formation simulations to study the completeness of the column density PDF. The clouds have been formed in the shocked layer between two supersonically converging streams. We have produced column 
density maps and PDFs of two different, but centrally condensed regions within the formed complex to study the shape and completeness of the PDF. In a next step, we calculated the value of the last closed contour in order to estimate the completeness limit of the N-PDFs. We have found qualitatively good agreement of the 
shape of the N-PDF with previous observational and numerical studies. We then showed that the last closed column-density contour moves towards higher column densities with decreasing size of the 
field of view. This means that, for too small sizes of the field of view, the N-PDF is only complete from the beginning of the power-law tail on towards the maximum column density. However, a sufficiently large field of view captures the log-normal part of the 
PDF. This was also confirmed by the analysis of the column density distribution of the whole complex.
Our results show that the N-PDFs of self-gravitating, turbulent molecular clouds are well described by a log-normal peak and a power-law tail. We conclude that a fully reliable observational study, which takes into account the completeness limit of the PDF must cover a sufficiently 
 large field of view, which might be difficult due to increasing contamination from neighbouring molecular clouds or a decreasing signal-to-noise ratio at the low column-density end of the PDF.

\section*{Acknowledgements}
The authors thank the anonymous referee for their insightful report, which helped to improve the quality of this study.
B.K.~and R.B.~acknowledge funding from the German Science Foundation (DFG) within the Priority Programm "The Physics of the ISM" (SPP 1573) via the grant BA 3706/3-2. R.B.~further 
acknowledges funding for this project from the DFG via the grants BA 3706/4-1, BA 3706/14-1 and BA 3706/15-1.
C.~F.~acknowledges funding provided by the Australian Research Council (Discovery Projects DP150104329 and DP170100603, and Future Fellowship FT180100495). B.K.~and C.F.~acknowledge funding via the Australia-Germany Joint Research Cooperation Scheme (UA-DAAD). 
C.F.~thanks for high-performance computing resources provided by the Leibniz Rechenzentrum 
and the Gauss Centre for Supercomputing (grants~pr32lo, pr48pi and GCS Large-scale project~10391), the Partnership 
for Advanced Computing in Europe (PRACE grant pr89mu), the Australian National Computational Infrastructure 
(grant~ek9), and the Pawsey Supercomputing Centre with funding from the Australian Government and the Government 
of Western Australia, in the framework of the National Computational Merit Allocation Scheme and the ANU Allocation 
Scheme. The simulations were run on HLRN--III under project grand hhp00022. The \textsc{flash} code was in 
part developed by the 
DOE--supported ASC/Alliance Center for Astrophysical Thermonuclear Flashes at the University of Chicago.
\bibliography{astro}
\bibliographystyle{mn2e}
\end{document}